\documentclass[english,aps,prper,reprint,showpacs,titlepage,longbibliography]{revtex4-2}   
\usepackage[compatibility=false]{caption} 

\usepackage{booktabs} 
\usepackage{tabularx} 
\usepackage{ragged2e} 
\usepackage{caption} 
\usepackage[table]{xcolor}
\usepackage{multirow}
\usepackage{graphicx}
\usepackage{rotating}
\usepackage{float}
\definecolor{lowcolor}{RGB}{255,252,210}
\definecolor{midcolor}{RGB}{215,255,215}
\definecolor{highcolor}{RGB}{245,220,230}
\usepackage[T1]{fontenc}	
\usepackage{geometry}
\usepackage{times}
\usepackage{hyperref}  
\hypersetup{colorlinks=true,urlcolor=blue,citecolor=blue,linkcolor=blue} 
\usepackage{array}
\usepackage{enumerate}
\usepackage{amsmath}
\usepackage{amssymb}
\usepackage{tikz}
\usepackage{titlesec}
\titleclass{\subsubsubsection}{straight}[\subsubsection]
\newcounter{subsubsubsection}[subsubsection]
\renewcommand{\thesubsubsubsection}
{\thesubsubsection.\arabic{subsubsubsection}}

\titleformat{\subsubsubsection}
{\normalfont\normalsize\bfseries}
{\thesubsubsubsection}{1em}{}

\titlespacing*{\subsubsubsection}
{0pt}{3.25ex plus 1ex minus .2ex}{1.5ex plus .2ex}

\usepackage{graphicx}
\usepackage{multirow}
\usepackage{tcolorbox}
\usepackage{ragged2e}
\usepackage{float}
\usepackage[utf8]{inputenc}

\usepackage[normalem]{ulem}
\usepackage{setspace} 

\begin{document}
\begin{titlepage}

\title{Integrative Thinking in Computational Physics Design Essays}

 \author{Sibatmika Babu}
 \affiliation{Department of Physics and Astronomy, Purdue University, 525 Northwestern Ave, West Lafayette, IN, 47907, U.S.A.} 

 \author{ Ravishankar Chatta Subramaniam}
 \affiliation{Department of Physics and Astronomy, Purdue University, West Lafayette, IN, 47907, U.S.A.} 

  \author{Carina M. Rebello}
 \affiliation{Department of Physics and Astronomy, Purdue University, West Lafayette, IN, 47907, U.S.A.} 

   \author{Jason W. Morphew}
 \affiliation{School of Engineering Education, Purdue University, West Lafayette, IN, 47907, U.S.A.}
 
 \author{N. Sanjay Rebello}
 \affiliation{Dept. of Physics \& Astronomy / Dept. of Curriculum \& Instruction, Purdue University, West Lafayette, IN, 47907, U.S.A.} 
   
  


\begin{abstract}
This study investigates integrative thinking in undergraduate computational physics design essays (CPDEs) within an engineering design-based introductory physics laboratory. Integrative thinking refers to the interaction of science, mathematics, computation, design, and metacognitive reflection during problem solving. Using the Ways of Thinking for Engineering Design-based Physics (WoT4EDP) framework, we examine how levels of integrative thinking can be characterized and manifested across teams with different levels of academic performance. We analyzed CPDEs from 14 student teams in a first-semester calculus-based physics course. Teams were grouped into high-, middle-, and low-performing categories based on members’ final examination performance, enabling comparison of integrative thinking across academic performance levels. CPDEs were analyzed using thick description and an analytic rubric characterizing three levels of integrative thinking: superficial, intermediate, and substantial. Findings illustrate variation in how students coordinate multiple ways of thinking within engineering design-based physics tasks and provide a structured lens for examining integrative thinking across levels of academic performance.

    \clearpage
  \end{abstract}

\maketitle
\end{titlepage}
\maketitle
\section{INTRODUCTION}
The American Association of Physics Teachers (AAPT) recommendations for undergraduate physics laboratories emphasize six essential learning goals: constructing knowledge, developing and using models, designing experiments, developing technical skills, analyzing and interpreting data, and communicating scientific results\cite{aapt2023lab}. Together, these practices support students’ development of scientific reasoning, critical thinking, collaboration, and ethical communication, providing a foundation for meaningful engagement in scientific inquiry. Engineering design (ED) activities further extend these goals by requiring students to engage in authentic, complex problem-solving processes that involve the integration of scientific principles, mathematical reasoning, design decisions, and metacognitive reflection \cite{capobianco2013,capobianco2018,NRC2012}.

To investigate how students engage in and connect these different forms of reasoning during engineering design tasks, the Ways of Thinking for Engineering Design-based Physics (WoT4EDP) framework was developed as a tool for characterizing student thinking within physics laboratory environments \cite{subramaniam2025stem,subramaniam2025design}. The framework identifies multiple ways of thinking that students may employ during ED activities, including science-based thinking (SBT), design-based thinking (DBT), mathematics-based thinking (MBT), computational thinking (CT), and metacognitive reflection (MER). By examining how these ways of thinking emerge and interact, the framework provides insight into the complexity of students’ problem-solving processes in physics-based design contexts.

Prior research on engineering design education has examined a variety of cognitive practices associated with successful design processes, including critical thinking, systems thinking, future-oriented thinking, strategic thinking, value-based thinking, and design thinking \cite{english2023stem,dalal2021ways}. Building upon this body of work, our previous research investigated multi-week engineering design activities in undergraduate physics laboratories and identified recurring STEM-related ways of thinking demonstrated by student teams, which informed the development of the WoT4EDP framework. Subsequent work has focused on applying this framework to student-generated engineering design tasks to better understand not only the presence of individual ways of thinking, but also how students connect and coordinate them during problem-solving.

The motivation for the present study is to examine \textit{Integrative Thinking} (IT) — the process through which students combine and coordinate multiple ways of thinking to develop solutions to engineering design challenges in physics classrooms \cite{subramaniam2025arxiv}. Although previous research using the WoT4EDP framework has characterized how students engage with individual elements of engineering design, including scientific concepts, mathematical reasoning, computational approaches, design decisions, and metacognitive reflection \cite{subramaniam2025stem}, limited research has examined how these elements interact with one another as students develop solutions. In particular, little is known about how students establish connections among different ways of thinking within computational physics design essays (CPDEs).

Understanding these connections is important because effective engineering design requires more than the application of isolated knowledge domains; it requires students to synthesize information across disciplines, evaluate design choices, and reflect on their reasoning. By examining the types of interactions and the levels of interaction among different ways of thinking, this study provides insight into how students integrate scientific, mathematical, computational, design, and reflective practices during engineering problem-solving. Such understanding can inform the development of instructional strategies and scaffolds that support students in developing more flexible, interconnected, and effective approaches to solving complex engineering challenges.

RQ1:
How can the levels of integrative thinking in engineering design-based physics tasks be characterized and operationalized?

RQ2:
How are different levels of integrative thinking manifested in students’ written explanations of engineering design-based physics problems?

\section{LITERATURE REVIEW}
Prior research has investigated how students engage in engineering design (ED)-based learning through multiweek design activities implemented in undergraduate physics laboratories \cite{subramaniam2025stem}. In this prior work, 14 student teams in a calculus-based physics course participated in instructor-assigned ED projects. Researchers analyzed students’ written design reports and team discussion transcripts to characterize students’ engagement in different forms of reasoning, including design-based thinking, science-based thinking, mathematical reasoning, and metacognitive reflection. While much prior research on engineering design education has emphasized challenges associated with students’ ability to connect scientific knowledge with design practices, often described as a design–science gap \cite{subramaniam2025stem}, this study highlighted the importance of examining design–science connections as evidence of students’ ability to apply physics concepts within authentic design contexts.

Building on this prior work \cite{subramaniam2025stem}, researchers developed the Ways of Thinking for Engineering Design-based Physics (WoT4EDP) framework to provide a structured approach for characterizing students’ reasoning processes during ED activities in physics contexts \cite{subramaniam2025design}. The framework was informed by national educational reform documents\cite{NRC2012,NRC2013NGSS}, contemporary physics education research (PER) literature, and prior studies on engineering design learning. Additionally, WoT4EDP was situated within broader STEM problem-solving research through comparisons with frameworks such as FAWTEER \cite{dalal2021ways} and Ways of Thinking in STEM-Based Problem Solving \cite{english2023stem}. These comparisons established the conceptual foundation of WoT4EDP while highlighting the need to further investigate how students coordinate multiple forms of reasoning during complex engineering problem-solving.

Subsequent research applied the WoT4EDP framework to student-generated engineering design tasks to examine how undergraduate students employed different ways of thinking while developing computational physics design solutions \cite{subramaniam2025arxiv}. This study analyzed written project reports and team discussion transcripts from 14 student teams to investigate students’ approaches to problem framing and their integration of computational tools, particularly python programming, within their design processes. The findings demonstrated that students engaged in multiple ways of thinking, including Science-based Thinking (SBT), Mathematics-based Thinking (MBT), Design-based Thinking (DBT), Metacognitive Reflections (MER) and Computational Thinking (CT) throughout the design process. DBT encompasses considerations related to the design process, including criteria and constraints, stakeholders, economic and safety considerations, design mechanisms and dimensions, limitations, and feasibility \cite{subramaniam2025stem,capobianco2018}. SBT captures students’ use of physics concepts, principles, and vocabulary, including assumptions and approximations used in applying fundamental principles such as energy, momentum, and angular momentum \cite{subramaniam2025stem,aapt2023lab,NRC2013NGSS,CollegeBoard2009}. MBT includes the identification and use of variables and constants, mathematical tools ranging from algebra and geometry to calculus, proportional and dimensional reasoning, qualitative reasoning, and the use of graphs and tables to identify patterns \cite{subramaniam2025stem, MAA2015CUPM, Tiles1984}. CT encompasses computational practices such as decomposition, pattern recognition, abstraction, algorithm design, automation, and debugging \cite{Shute2017CT,LopezParra2023CT}. Finally, MER captures students’ reflection on their design process, including the need for iteration and reflection on the scientific concepts relevant to their design decisions \cite{subramaniam2025stem, Dori2018MetacognitionSTEM,Hacker2003Metacognition, NRC2000HowPeopleLearn}. Together, these five ways of thinking provide the conceptual foundation for examining how students engage with different forms of reasoning during physics-based engineering design.

While our previous studies \cite{subramaniam2025design, subramaniam2025stem} have provided insight into the presence and characteristics of individual ways of thinking within ED tasks, limited research has examined how these ways of thinking interact and become integrated as students develop solutions. Specifically, little is known about the depth and nature of connections students establish among scientific concepts, mathematical reasoning, design decisions, computational approaches, and metacognitive reflection.

The present study builds on this prior work by focusing on \textit{integrative thinking} (IT). Although IT has been discussed across multiple disciplines, there is limited research that clearly defines and measures IT within physics-based engineering design contexts. Recent work in education and business suggests that IT involves the ability to perceive holistic and interconnected systems, synthesize knowledge across multiple disciplines and perspectives, reconcile competing ideas, and engage in reflection and metacognition \cite{Brody2026}. Building on these perspectives, we operationally define \textit{integrative thinking} as the coexistence and interaction of two or more ways of thinking—such as science-based thinking (SBT), design-based thinking (DBT), mathematics-based thinking (MBT), computational thinking (CT), and metacognitive reflection (MER), while solving physics-based engineering design problems.

Within this study, we examine IT in Computational Physics Design Essays (CPDEs), an adaptation of computational essays introduced by Odden et al. \cite{odden2023using,odden2025physics}. In CPDEs, students use computational tools (typically \textit{Python}) to investigate real-world challenges involving physics principles and communicate their solutions through integrated artifacts, including written explanations, code, equations, figures, and tables. Although CPDEs provide opportunities for students to combine multiple forms of reasoning, less is known about how evidence of IT emerges within these student-generated artifacts.

This  study addresses this gap by examining how students demonstrate IT within CPDEs. Specifically, we focus on how students integrate science-based thinking, mathematics-based thinking, design-based thinking, computational thinking, and metacognitive reflection to construct and communicate solutions to physics-based engineering design challenges. By investigating the depth and nature of these connections, this study extends prior WoT4EDP research by moving beyond identifying individual ways of thinking toward understanding how students coordinate and integrate these ways of thinking during complex problem-solving.

\section{METHODS}

\subsection{Context of Study and Participants}

The participants in this study were enrolled in a first-semester calculus-based undergraduate physics course at a large Midwestern U.S. land-grant university. Over 80\% of students enrolled in the course were engineering majors, with the remaining students pursuing science-related disciplines. As part of the laboratory curriculum, students engaged in engineering design (ED)-based activities that integrated computational modeling, physics reasoning, and design practices. 

The CPDE weekly journal guided students through the development of their engineering design problem over several weeks. Students initially brainstormed and scoped potential problems before developing a problem statement. They then refined their problem, identified relevant stakeholders, criteria, constraints, and physics principles, and began developing an initial solution. In subsequent weeks, students continued to develop and refine their solutions through computational modeling, physics reasoning, and iterative design. Students also engaged in team discussions focused on their problem and the application of physics principles.

\begin{table}[htbp]

\captionsetup{justification=raggedright,singlelinecheck=false}
\caption{Weekly Engineering Design Activities}
\label{table:ED_activities}
\scriptsize
\renewcommand{\arraystretch}{0.85}
\setlength{\tabcolsep}{3pt}
\begin{ruledtabular}
\noindent
\begin{tabular}{ll}
\textbf{\parbox[t]{0.7cm}{Week}} &
\textbf{\parbox[t]{6.8cm}{ED-task activities}} \\

10 &
\parbox[t]{6.8cm}{\RaggedRight Students receive basic information about the student-generated ED problem and individually engage in problem scoping by brainstorming multiple ideas and narrowing their choice to one problem. Student teams then engage in a free-flowing team discussion considering stakeholders, criteria, constraints, and relevant physics principles.} \\

11 &
\parbox[t]{6.8cm}{\RaggedRight Students receive a rubric and a Jupyter Notebook template for their ED problem. An example ED problem is provided. Students revisit and refine their problem statement if needed, break it into simpler parts, and begin working on the Jupyter Notebook.} \\

12 &
\parbox[t]{6.8cm}{\RaggedRight Students continue working on their ED problem and engage in a free-flowing team discussion focused on their specific problem statement and how they intend to apply physics principles in their solution.} \\

13 &
\parbox[t]{6.8cm}{\RaggedRight Students continue refining and developing their solutions.} \\

14 &
\parbox[t]{6.8cm}{\RaggedRight The entire class time is dedicated to the ED task. Student teams finalize their problem statement, design thinking, science thinking, mathematical thinking, first attempt, first iteration, and optional second iteration. Teams also discuss their iterative progression, refinements, and application of physics principles.}

\end{tabular}
\end{ruledtabular}
\end{table}

The dataset for this study consisted of artifacts from all 14 student teams enrolled in one laboratory section taught by one of the co-authors. Each team consisted of three students. Data sources included students' final Computational Physics Design Essay (CPDE) reports and transcripts of team discussions collected from weeks 10 through 14 of a 16-week semester. These artifacts provided opportunities to examine how students connected multiple ways of thinking while engaging in engineering design-based physics tasks.

The final CPDE report consolidated this work into a structured account of the design process. The report included the problem statement, initial solution and its limitations, first iteration and its limitations, and an optional second iteration. The second iteration was not required and was included when students further refined their solution. The report concluded with the final solution and conclusions. Thus, the CPDE provided a record of students' progression from problem definition through initial solution development, iteration, and final conclusions. Table~\ref{tab:cpde_topic} presents the problems addressed by each team. 

\begin{table}[h]
\centering
\caption{Problems addressed by the 14 teams in their CPDEs.}
\label{tab:cpde_topic}
\begin{tabular}{ll}
\hline
\textbf{Team} & \textbf{CPDE Problem} \\
\hline
T01 & Optimizing Conditions for a Baseball Home Run \\
T02 & Optimizing Autonomous Golf Cart Travel \\
T03 & Optimizing Mortar Projectile Range \\
T04 & Maximizing Catapult Projectile Range \\
T05 & Designing a Hookean Bow for Archery \\
T06 & Maximum Cornering Speed of a Formula 1 Car \\
T07 & Efficient and Safe Drone Delivery \\
T08 & Hammer Mass and Nail-Driving Force \\
T09 & Safe Toy Catapult Design \\
T10 & Force Required to Stop a Bullet \\
T11 & Maximum Safe Mass for Bungee Jumping \\
T12 & Skater and Wheel Speed on a Ramp \\
T13 & Maximum Height and Range of a Thrown Axe \\
T14 & Comparing Yo-Yo Designs for Angular Speed \\
\hline
\end{tabular}
\end{table}

\subsection{Analytical Framework}

Our analysis was guided by the Ways of Thinking for Engineering Design-based Physics (WoT4EDP) framework \cite{subramaniam2025design}, which describes the ways students draw upon and connect different ways of thinking during engineering design activities. The WoT4EDP framework identifies five ways of thinking in physics-based engineering design: Design-Based Thinking (DBT), Science-Based Thinking (SBT), Mathematics-Based Thinking (MBT), Computational Thinking (CT), and Metacognitive Reflection (MER). These ways of thinking involve design processes, physics reasoning, mathematical reasoning, computational tools, and reflection on problem-solving, respectively. The framework emphasizes that multiple ways of thinking can coexist and interact during engineering design. Specifically, we examined connections among Science-Based Thinking (SBT), Design-Based Thinking (DBT), Math-Based Thinking (MBT), Computational Thinking (CT), and Metacognitive Reflection (MER).
\begin{figure}[hbt]
    \centering
    \includegraphics[width=1\linewidth]{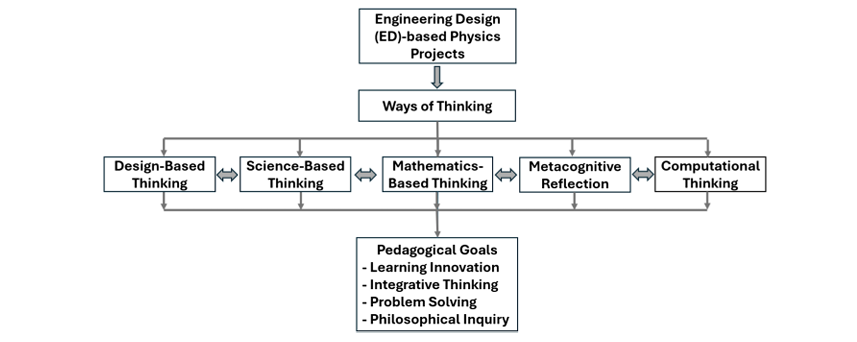}
    \caption{Framework illustrating the different ways of thinking in physics-based engineering design. Adapted from \cite{subramaniam2025stem}}
    \label{fig:placeholder}
\end{figure}
To characterize the depth of students' integrative thinking (IT), we developed an analytic rubric through an iterative analysis of student artifacts. In order to develop this rubric\ref{tab:IT_examples}, the CPDEs were first examined using the WoT4EDP framework to identify relevant ways of thinking and the connections among them. Through repeated examination of student responses, we refined the criteria describing the nature and depth of the connections between the different ways of thinking, and developed three levels of integrative thinking. The initial rubric was independently applied by three raters to a subset of three CPDEs from the 14 teams to evaluate the clarity and applicability of the rubric. The raters then discussed their coding and reached consensus. We used these discussions to refine the rubric definitions and establish a consistent interpretation of the three levels of integrative thinking. The finalized rubric was subsequently applied to all 14 team CPDEs to characterize the depth of students' integrative thinking connections.

\subsection{Quantitative Analysis of Integrative Thinking}

To examine the variation in integrative thinking across levels of academic performance, students were first ranked according to their final examination scores. High-performing students were those whose final exam scores were more than one standard deviation (+10\%) above the class average (78\%), low-performing students were those whose scores were more than one standard deviation (-10\%) below the class average, and mid-performing students were those whose scores fell between these ranges. Teams were then characterized based on the average final examination score of their members. Based on this ranking, teams were teamed as follows: Teams 1, 4, 5, and 12 were classified as high-performing; Teams 3, 7, 9, 11, and 14 were classified as mid-performing; and Teams 2, 6, 8, 10, and 13 were classified as low-performing. Table~\ref{tab:team_perf} summarizes individual student scores (rank), team averages, and performance team assignments.

\begin{table}[h]
\centering
\caption{Final exam scores of students, team averages, and performance team classification. Each score is classified as Low- (L), Mid- (M), or High- (H) performing.}
\label{tab:team_perf}
\begin{tabular}{llllll}
\toprule
Team &  Student 1 & Student 2 & Student 3 & Team Avg. \\
\midrule
T01 & 80\% (M) & 100\% (H)  & 97\% (H) & 92.33\% (H) \\
T02 & 67\% (L) & 90\% (H) & 73\% (M) & 76.67\% (L) \\
T03 & 77\% (M) & 80\% (M) & 83\% (M) & 80.00\% (M) \\
T04 & 97\% (H) & 77\% (M) & 93\% H) & 89.00\% (H) \\
T05 & 87\% (M) & 93\% (H) & 87\% (M) & 89.00\% (H) \\
T06 & 53\% (L) & 100\% (H) & 77\% (M)& 76.67\% (L) \\
T07 & 77\% (M) & 100\% (H) & 100\% (H) & 84.67\% (M) \\
T08 & 80\% (M) & 70\% (M) & 73\% (M) & 74.33\% (L) \\
T09 & 83\% (M) & 80\% (M) & 87\% (M) & 83.33\% (M) \\
T10 & 87\% (M)  & 80\% (M) & 43\% (L) & 70.00\% (L) \\
T11 & 83\% (M) & 90\% (H) & 90\% (M) & 84.33\% (M) \\
T12 & 90\% (H) & 93\% (H) & 87\% (M) & 90.00\% (H) \\
T13 & 53\% (L) & 100\% (H) & 67\% (L) & 73.33\% (L) \\
T14 & 93\% (H) & 63\% (L) & 87\% (M) & 81.00\% (M) \\
\bottomrule
\end{tabular}
\end{table}

To identify instances of integrative thinking within the Computational Physics Design Essays (CPDEs), we first applied the Ways of Thinking for Engineering Design Processes (WoT4EDP) framework to all 14 team reports. This framework was used to identify the presence of different STEM-related ways of thinking, including Science-Based Thinking (SBT), Design-Based Thinking (DBT), Mathematics-Based Thinking (MBT), Computational Thinking (CT), and Metacognitive Reflection (MER). We then examined how these ways of thinking coexisted within students' explanations. Specifically, we identified instances where two or more ways of thinking were connected within the same paragraph, sentence, or closely related statements to explain or justify an idea, design decision, or interpretation of results. These connections were considered evidence of integrative thinking.

Using the finalized rubric, each identified instance of integration between ways of thinking was coded and categorized into one of three levels: Level 1 (Superficial), Level 2 (Intermediate), or Level 3 (Substantial). 

Superficial IT was defined as instances where students demonstrated connections between two or more ways of thinking in an implicit or surface-level manner without elaboration or justification. For example:\textit{"Throughout the process we determined that the optimal way to work on this problem will be to find the distance of the bean bag when the acted on by known constants, instead of trying to find the constant K for the spring force. This is due to the extreme complexity of this process."}

The team demonstrated superficial IT in MBT-MER connection by referring to the mathematical parameter K for spring force and reflecting on the complexity of determining it. However, they provided little explanation on why this approach was considered too complex. 

Intermediate IT represented connections supported by evidence, reasoning, or partial justification, demonstrating a clearer relationship between ways of thinking. For example:\textit{"Our main limitation are primarily that we have less control over how the pitcher pitches. To simlplify our problem, we will be assuming and standardizing certain conditions such
as the mass of the ball, inital speed of the ball, the contact time of the bat and the ball."} 

The team demonstrated intermediate IT in the SBT-DBT connection by identifying the limitation of their problem because of which they propose using standardize conditions. Here, they are making assumption for practicality of the problem there by showing connection between understanding physical factor affecting the design decision but aren't explaining the reason behind selecting these specific condition. 

Substantial IT was characterized by well-developed connections in which students meaningfully integrated ways of thinking through explicit reasoning, justification, and coordination of ideas. For example: \textit{"The size and mass of the projectile will be constant, and so will the force of gravity. Constants: Mass of projectile: 40 g. Catapult arm: 0.30 m Spring Constant:100 N/m  "}.

The team demonstrated substantial IT in the SBT–MBT connection by explicitly linking physics concepts, including spring force and projectile motion, with specific mathematical quantities. By identifying the projectile mass (40 g) and spring constant (100 N/m) as constants, they connected the physical properties of the system to mathematical representations relevant to their design. 

The coded data were then analyzed to determine the overall frequency and distribution of integrative thinking across the 14 teams. We first examined the frequency of connections between different ways of thinking across all teams to identify common patterns of integration. We then analyzed the distribution of integrative thinking across the three rubric levels to understand the depth at which teams demonstrated integration among STEM ways of thinking.

To compare integrative thinking across high-, mid-, and low-performing teams, the frequency of each type and level of integrative thinking was averaged within each performance team. The frequency of connections was used as a descriptive measure of how often connections among different ways of thinking were made visible in students’ written work, allowing comparisons of patterns across performance teams. Specifically, team-level counts of integrative thinking connections were averaged across all teams classified as high-performing, mid-performing, and low-performing. This approach allowed us to compare patterns of integration while accounting for differences in the number of teams within each performance category.

\subsection{Qualitative Analysis of Integrative Thinking}

To examine how integrative thinking was constructed within student work, we selected one representative team from each performance category (high, mid, and low) for detailed qualitative analysis. Selection of cases was guided by the goal of capturing variation in the extent and expression of integrative thinking across teams.

To examine how integrative thinking was constructed within student work, one representative team from each performance category was selected for in-depth qualitative analysis. Random selection was used to avoid introducing additional selection criteria or researcher bias in choosing teams within each performance category. Teams 1, 9, and 10 were selected as representative cases for the high-, mid-, and low-performing teams, respectively.

The selected cases were analyzed using a "thick description" approach that combined researcher interpretation with excerpts from students' written artifacts to preserve the context in which integrative thinking was expressed \cite{Brink1987, peel2020beginners}. {As Ponterotto says:
\textit{“A thickly described Discussions section of
a qualitative report successfully merges the
participants’ lived experiences with the researcher’s interpretations of these experiences…”}
(\cite{Ponterotto2006}, p. 547)}. While the analytic rubric provided a systematic approach for identifying and categorizing the depth of students’ integrative thinking (IT), thick description was used to provide detailed, context-sensitive interpretations of how these connections were constructed and expressed in students’ design work. This approach was warranted because assigning a depth level alone neither fully captures the nature of students’ reasoning nor how different ways of thinking (WoT4EDP) are connected within the design process. The thick description allowed us to examine students’ statements in relation to the surrounding explanations, design decisions, mathematical reasoning, computational work, and reflections in their CPDEs. This contextual interpretation supported a more nuanced understanding of why an IT connection represented a particular level of depth and how students developed and applied these connections while solving the engineering design problem. The WoT4EDP framework served as the conceptual lens for interpreting students’ thinking and the analytic rubric provided the basis for characterizing its depth.

This combination of quantitative comparison and qualitative case analysis enabled us to examine both the distribution of integrative thinking across teams and the ways in which students enacted integrative thinking during engineering design-based physics activities.

\section{FINDINGS}
\subsection{RQ1: Characterizing and Operationalizing the Depth of Integrative Thinking}
The qualitative coding of the 14 team CPDE reports, guided by the WoT4EDP framework and the finalized analytic rubric, resulted in an operational framework for characterizing the depth of integrative thinking in students’ written work. The analysis distinguished three levels of integrative thinking: superficial (Level 1), intermediate (Level 2), and substantial (Level 3).

As illustrated in Table~\ref{tab:IT_examples}, the three levels distinguish the extent to which students meaningfully connect different ways of thinking within their explanations. At the superficial level, multiple ways of thinking are present, but the relationship between them is minimally developed. At the intermediate level, students establish a meaningful connection between ways of thinking and provide some supporting evidence, reasoning, or justification. At the substantial level, students develop and explicitly justify the relationship between different ways of thinking, demonstrating how their integration informs a design decision, interpretation, or explanation.
Table~\ref{tab:IT_examples} presents the resulting analytic rubric and representative examples from students' CPDE reports.

This operationalization therefore characterizes integrative thinking not only by the presence of multiple ways of thinking, but also by the depth of the connections constructed among them. The resulting framework provided the basis for examining variation in the frequency and depth of integrative thinking across the 14 teams.
\begin{table*}
\centering
\caption{Examples of Integrative Thinking at Different Levels of Integration}
\label{tab:IT_examples}

\scriptsize
\setlength{\tabcolsep}{3pt}
\renewcommand{\arraystretch}{0.85}

\begin{tabular*}{\textwidth}{@{\extracolsep{\fill}}c c c c@{}}
\hline

\textbf{IT} &
\textbf{Level 1 - Superficial} &
\textbf{Level 2 - Intermediate} &
\textbf{Level 3 - Substantial} \\

\hline

\textbf{SBT - MBT} &
\parbox[t]{0.30\textwidth}{\RaggedRight Identifies the relevant physics term or concept and mentions basic mathematical elements such as calculations or variables \\ \textit{"Going into the momentum equation with two unknowns is a major limitation."}} &
\parbox[t]{0.30\textwidth}{\RaggedRight Identifies and explains the physics concept and uses specific mathematical terms, including what is being calculated or the name of the equation.\\ \textit{"In this problem, we will focus on the batter's perspective and calculate the exact time, position, and technique that they should hit the ball in order to get the best outcome."}} &
\parbox[t]{0.30\textwidth}{\RaggedRight Discusses the physics concept in detail and clearly states the equation, mathematical values used in the calculation, or equation.\\ \textit{"The work required by the engine is at least 22.78 kJ. Of course, this value is by the work-energy principle, which we have established is not the most ideal for our solution, but gives us a minimum."}}\\
\vspace{0.15 cm}

\textbf{SBT - DBT} &
\parbox[t]{0.30\textwidth}{\RaggedRight Identifies physics-related terms, concepts, or principles and states the problem/idea/solution.\\ \textit{"ModernGolfersUnlimited is an Indiana based company that produces golf carts. What makes them unique is that they have been working on creating golf carts that are self driving, and able to move from hole to hole following the most optimal and safest path."}} &
\parbox[t]{0.30\textwidth}{\RaggedRight Explains the physics idea and expands it by describing criteria, constraints, stakeholder or aspects of iteration.\\ \textit{"Our main limitation are primarily that we have less control over how the pitcher pitches. To simplify our problem, we will be assuming and standardizing certain conditions such as the mass of the ball, initial speed of the ball, and the contact time of the bat and the ball."}} &
\parbox[t]{0.30\textwidth}{\RaggedRight Discusses the physics concept thoroughly and presents a clear, concrete design statement or idea.\\ \textit{"Therefore, a primary concern when designing a catapult is maximizing its effective range to reduce these risks as much as possible. To this effect, the question of this essay is what are the optimal launch angle of a projectile of a catapult to maximize the distance the projectile will travel?"}} \\ \vspace{0.15 cm}

\textbf{SBT - MER} &
\parbox[t]{0.30\textwidth}{\RaggedRight Identifies the physics concept and reflects on it vaguely.\\ \textit{"Using our physics knowledge that we have acquired from this class, we can calculate speed of pitch and how much spin a pitcher puts on a ball in order to predict the outcome of the ball's trajectory."}} &
\parbox[t]{0.30\textwidth}{\RaggedRight Explains the concept and explain on where they have previously encountered it.\\ \textit{"Roughly the same specification existed, but the main differences was that we did not include a final speed of the arrow in the specifications, which caused some problems, hence why this was only the first solution."}} &
\parbox[t]{0.30\textwidth}{\RaggedRight Discusses the physics concept and reflects on how it helps in solving their problem.\\ \textit{"The relevant principles and concepts for this problem are the energy principle ... The energy principle applies since energy is neither created nor destroyed but the axe's energy is converted to different forms in the process."}}\\ \vspace{0.15 cm}

\textbf{SBT - CT} &
\parbox[t]{0.30\textwidth}{\RaggedRight Identifies the physics concept and uses terms such as “codes” or “Python.”\\ \textit{"For both parts, we iterate through a range of values and then plot the results of our ``simulation''."}} &
\parbox[t]{0.30\textwidth}{\RaggedRight Explains the physics concept and describes the situation in which code is used. \\ \textit{"In our code we find the rotational speed from the net angular momentum with the given perpendicular distance."}} &
\parbox[t]{0.30\textwidth}{\RaggedRight Discusses the physics concept and explains how the code is being used to solve a specific physics component or includes explicit mentions of coding.\\ \textit{"
\texttt{rope\_stretch = rope\_length * stretch}\\
\texttt{spring\_constant = catapult\_tension / rope\_stretch}\\
\texttt{spring\_energy = 0.5 * spring\_constant * rope\_stretch**2}\\
\texttt{print(spring\_energy)}"}
} \\\vspace{0.15 cm}

\textbf{MBT - DBT} &
\parbox[t]{0.30\textwidth}{\RaggedRight Uses mathematical terms (e.g., calculation, variables) to state the problem, solution, or criteria. \\ \textit{"We had a hard time making calculations because there were too many unknowns. We also wanted to allow the user more control over the calculations by including user inputs."}} &
\parbox[t]{0.30\textwidth}{\RaggedRight Explicitly uses specific mathematical terms, formulas, or equation names to propose or refine a solution, including iteration. \\ \textit{In this problem, we are assuming that pure rolling is going on so we are using the equation $\omega=v_{cm}/R$, where $R$ is the radius of the wheels and $v_{cm}$ is equal to the final velocity."}} &
\parbox[t]{0.30\textwidth}{\RaggedRight Uses numerical values of given variables and connects them to the solution while stating the problem or criteria.\\ \textit{"However, we can make a function to plot the distance of every launch angle between 0 and 90 degrees ... we can solve for the vertex of the distance function to find the ideal launch angle."}} \\\vspace{0.15 cm}

\textbf{MBT - MER} &
\parbox[t]{0.30\textwidth}{\RaggedRight Uses mathematical terms such as calculation or variables and reflects vaguely on how they help.\\ \textit{"After doing a bunch of research, we found the constants and equations we need to solve this problem."}} &
\parbox[t]{0.30\textwidth}{\RaggedRight Explicitly uses specific mathematical terms, formulas, or equations and explain on where they were previously encountered or how they relate to the problem or reflects on their result based on calculation. \\ \textit{One of the biggest things we lacked in the early labs was numbers. That is why we tried to provide as much mathematical backing as we could into everything we showed in this CPE."}} &
\parbox[t]{0.30\textwidth}{\RaggedRight Uses numerical values and explains why these values were chosen and why a particular equation is being used. \\ \textit{"The main equations we used to use for this problem are found in the Energy principle equation of change in Energy = Q + W, as well as the main Momentum principle equation of change $p=p_i+F_{\mathrm{net}}(t)$."}} \\\vspace{0.15 cm}

\textbf{MBT - CT} &
\parbox[t]{0.30\textwidth}{\RaggedRight Uses mathematical terms like calculation or variables along with informal references to code or Python.\\ \textit{"This code cell is to define any constants that will be present for all runs, regardless of what is inputted."}} &
\parbox[t]{0.30\textwidth}{\RaggedRight Uses specific terms, formulas, or equations and identifies where code is being applied.  \\ \textit{"In Python we can only solve systems of first order ODEs, so defining $v_x=\dot{x}$ and $v_y=\dot{y}$ we get a system of 4 coupled first order ODEs."}} &
\parbox[t]{0.30\textwidth}{\RaggedRight Defines values for variables and explains how code is being used or includes explicit code references.\\ \textit{"
\texttt{R = (2*KE*math.sin(2*theta\_small))/(m*g)}\\
\texttt{r = R/1000}\\
\texttt{v\_i = math.sqrt((2*KE\_trans)/m)}\\
\texttt{w = v\_i*.03048}"}
} \\\vspace{0.15 cm}

\textbf{DBT - MER} &
\parbox[t]{0.30\textwidth}{\RaggedRight Reflects on the practicality of the problem or solution.\\ \textit{"As stated above, the way that we handled the impact angle with respect to total momentum was originally incorrect, and this continued into iteration one."}} &
\parbox[t]{0.30\textwidth}{\RaggedRight Reflects in detail on the criteria and limitations, explaining why particular decisions were made. \\ \textit{"The limits of this iteration is that we will need to know what the angle and the translational kinetic energy are. Without either, we are not able to figure out what the different components are which we use to find the solution."}} &
\parbox[t]{0.30\textwidth}{\RaggedRight Reflects on why the chosen solution works and identifies conditions under which it will or will not work.\\ \textit{"The limitation of this solution is that our calculation can only be a finite point. This solution can allow us to compare distinct launch angles, but not to find the ideal launch angle."}} \\\vspace{0.15 cm}

\textbf{DBT - CT} &
\parbox[t]{0.30\textwidth}{\RaggedRight Identifies whether the problem statement includes any reference to code. \\ \textit{"The limitations of our first iteration ... we were not showing much data in the first place. We decided that we really needed a graph in order to better show the user..."}} &
\parbox[t]{0.30\textwidth}{\RaggedRight Indicates how code is involved in generating or refining the solution, explaining where coding is used. \\ \textit{"The kinetic energy from gunpowder will be entered by the user, as well as the angle used. We will create a code that will be able to take each of the metrics listed above and output the distance."}} &
\parbox[t]{0.30\textwidth}{\RaggedRight The design explicitly includes written code or explains how the code will be applied in the solution.\\ \textit{"Once all of these variables are plugged in, we solve for $\omega$ .... Once time and velocity are found we can find the maximum height of the projectile using the position update formula."}} \\\vspace{0.15 cm}

\textbf{MER - CT} &
\parbox[t]{0.30\textwidth}{\RaggedRight Reflects on their use of Python, only mentioning that they used code without explaining why or how it contributed to their thinking or solution. \\ \textit{"Ultimately, we decided to use the momentum principle to solve for the time taken from hole to hole, and the energy principle to solve for the work the engine needed to exert. The first part is done on paper, whereas the second is done in the coding cells below."}} &
\parbox[t]{0.30\textwidth}{\RaggedRight Explains how Python supported part of their reasoning—such as identifying where the code was applied, what it helped them calculate or visualize, or how it connected to their understanding of the problem.\\ \textit{"In our code we find the rotational speed from the net angular momentum with the given perpendicular distance. From this we can then use energy principle ... we can get a launch velocity and then momentum to simulate the trajectory following the impact."}} &
\parbox[t]{0.30\textwidth}{\RaggedRight Reflect on how Python coding shaped their scientific and design reasoning.  \\ \textit{"This is the calculation cell. The purpose of this is to find what the distance will be in kilometers that the shell will travel. Then, we found what the initial velocity is from that the $KE_{\mathrm{trans}}$ value that was inputted. After this, we calculated what the rotational kinetic energy will be."}} \\

\hline
\end{tabular*}
\end{table*}
\subsection{RQ2: Quantitative Analysis of Integrative Thinking Connections}
To examine how integrative thinking varied across teams with different levels of academic performance, we analyzed the frequency and depth of IT connections identified in the CPDE reports. A connection was identified when two or more ways of thinking were used together within a meaningful explanatory unit, such as the same sentence or closely connected statements, to explain, justify, or interpret an idea, design decision, or result. After applying the finalized integrative thinking rubric to all 14 teams, each connection between two ways of thinking was coded according to its level of integration: Level 1 (Superficial), Level 2 (Intermediate), and Level 3 (Substantial). 

Because the analysis involved comparisons at both the team and performance-team levels, we first examined the distribution of IT connections across individual teams. We then calculated average IT connections within high-, mid-, and low-performing teams to compare broader patterns of integrative thinking.

\begin{table*}[t]
\centering
\caption{Team Academic Performance and Integrative Thinking Connections}
\label{tab:team_it}

\scriptsize
\setlength{\tabcolsep}{5pt}
\renewcommand{\arraystretch}{0.95}

\begin{tabular}{c l c c c c c c}
\hline
\textbf{Team} &
\textbf{Performance} &
\textbf{Average} &
\textbf{Total IT} &
\textbf{Substantial} &
\textbf{Intermediate} &
\textbf{Superficial} &
\textbf{Page} \\
\hline

\cellcolor{highcolor}\multirow{3}{*}{\textcolor{red}{\textbf{T01}}}
& \cellcolor{highcolor}\textcolor{red}{\textbf{Mid}} & & & & & & \\
& \cellcolor{highcolor}\textcolor{red}{\textbf{High}} &
\cellcolor{highcolor}\textcolor{red}{\textbf{92.33}} &
\cellcolor{highcolor}\textcolor{red}{\textbf{70}} &
\cellcolor{highcolor}\textcolor{red}{\textbf{27}} &
\cellcolor{highcolor}\textcolor{red}{\textbf{21}} &
\cellcolor{highcolor}\textcolor{red}{\textbf{22}} &
\cellcolor{highcolor}\textcolor{red}{\textbf{12}} \\
& \cellcolor{highcolor}\textcolor{red}{\textbf{High}} & & & & & & \\
\hline

\cellcolor{lowcolor}\multirow{3}{*}{\textbf{T02}}
& \cellcolor{lowcolor}Low & & & & & & \\
& \cellcolor{lowcolor}High &
\cellcolor{lowcolor}76.66 &
\cellcolor{lowcolor}49 &
\cellcolor{lowcolor}28 &
\cellcolor{lowcolor}12 &
\cellcolor{lowcolor}9 &
\cellcolor{lowcolor}5 \\
& \cellcolor{lowcolor}Mid & & & & & & \\
\hline

\cellcolor{midcolor}\multirow{3}{*}{\textbf{T03}}
& \cellcolor{midcolor}Mid & & & & & & \\
& \cellcolor{midcolor}Mid &
\cellcolor{midcolor}80 &
\cellcolor{midcolor}63 &
\cellcolor{midcolor}21 &
\cellcolor{midcolor}18 &
\cellcolor{midcolor}24 &
\cellcolor{midcolor}4 \\
& \cellcolor{midcolor}Mid & & & & & & \\
\hline

\cellcolor{highcolor}\multirow{3}{*}{\textbf{T04}}
& \cellcolor{highcolor}High & & & & & & \\
& \cellcolor{highcolor}Mid &
\cellcolor{highcolor}89 &
\cellcolor{highcolor}74 &
\cellcolor{highcolor}36 &
\cellcolor{highcolor}12 &
\cellcolor{highcolor}26 &
\cellcolor{highcolor}10 \\
& \cellcolor{highcolor}High & & & & & & \\
\hline

\cellcolor{highcolor}\multirow{3}{*}{\textbf{T05}}
& \cellcolor{highcolor}Mid & & & & & & \\
& \cellcolor{highcolor}High &
\cellcolor{highcolor}89 &
\cellcolor{highcolor}41 &
\cellcolor{highcolor}23 &
\cellcolor{highcolor}7 &
\cellcolor{highcolor}11 &
\cellcolor{highcolor}4 \\
& \cellcolor{highcolor}Mid & & & & & & \\
\hline

\cellcolor{lowcolor}\multirow{3}{*}{\textbf{T06}}
& \cellcolor{lowcolor}Low & & & & & & \\
& \cellcolor{lowcolor}High &
\cellcolor{lowcolor}76.66 &
\cellcolor{lowcolor}32 &
\cellcolor{lowcolor}22 &
\cellcolor{lowcolor}7 &
\cellcolor{lowcolor}3 &
\cellcolor{lowcolor}4 \\
& \cellcolor{lowcolor}Mid & & & & & & \\
\hline

\cellcolor{midcolor}\multirow{3}{*}{\textbf{T07}}
& \cellcolor{midcolor}Mid & & & & & & \\
& \cellcolor{midcolor}High &
\cellcolor{midcolor}84.67 &
\cellcolor{midcolor}28 &
\cellcolor{midcolor}14 &
\cellcolor{midcolor}10 &
\cellcolor{midcolor}4 &
\cellcolor{midcolor}4 \\
& \cellcolor{midcolor}High & & & & & & \\
\hline

\cellcolor{lowcolor}\multirow{3}{*}{\textbf{T08}}
& \cellcolor{lowcolor}Mid & & & & & & \\
& \cellcolor{lowcolor}Mid &
\cellcolor{lowcolor}74.33 &
\cellcolor{lowcolor}49 &
\cellcolor{lowcolor}17 &
\cellcolor{lowcolor}6 &
\cellcolor{lowcolor}26 &
\cellcolor{lowcolor}6 \\
& \cellcolor{lowcolor}Mid & & & & & & \\
\hline

\cellcolor{midcolor}\multirow{3}{*}{\textcolor{red}{\textbf{T09}}}
& \cellcolor{midcolor}\textcolor{red}{\textbf{Mid}} & & & & & & \\
& \cellcolor{midcolor}\textcolor{red}{\textbf{Mid}} &
\cellcolor{midcolor}\textcolor{red}{\textbf{83.33}} &
\cellcolor{midcolor}\textcolor{red}{\textbf{36}} &
\cellcolor{midcolor}\textcolor{red}{\textbf{29}} &
\cellcolor{midcolor}\textcolor{red}{\textbf{4}} &
\cellcolor{midcolor}\textcolor{red}{\textbf{3}} &
\cellcolor{midcolor}\textcolor{red}{\textbf{7}} \\
& \cellcolor{midcolor}\textcolor{red}{\textbf{Mid}} & & & & & & \\
\hline

\cellcolor{lowcolor}\multirow{3}{*}{\textcolor{red}{\textbf{T10}}}
& \cellcolor{lowcolor}\textcolor{red}{\textbf{Mid}} & & & & & & \\
& \cellcolor{lowcolor}\textcolor{red}{\textbf{Mid}} &
\cellcolor{lowcolor}\textcolor{red}{\textbf{70}} &
\cellcolor{lowcolor}\textcolor{red}{\textbf{43}} &
\cellcolor{lowcolor}\textcolor{red}{\textbf{26}} &
\cellcolor{lowcolor}\textcolor{red}{\textbf{7}} &
\cellcolor{lowcolor}\textcolor{red}{\textbf{10}} &
\cellcolor{lowcolor}\textcolor{red}{\textbf{4}} \\
& \cellcolor{lowcolor}\textcolor{red}{\textbf{Low}} & & & & & & \\
\hline

\cellcolor{midcolor}\multirow{3}{*}{\textbf{T11}}
& \cellcolor{midcolor}Mid & & & & & & \\
& \cellcolor{midcolor}High &
\cellcolor{midcolor}84.33 &
\cellcolor{midcolor}32 &
\cellcolor{midcolor}15 &
\cellcolor{midcolor}10 &
\cellcolor{midcolor}7 &
\cellcolor{midcolor}3 \\
& \cellcolor{midcolor}Mid & & & & & & \\
\hline

\cellcolor{highcolor}\multirow{3}{*}{\textbf{T12}}
& \cellcolor{highcolor}High & & & & & & \\
& \cellcolor{highcolor}High &
\cellcolor{highcolor}90 &
\cellcolor{highcolor}35 &
\cellcolor{highcolor}18 &
\cellcolor{highcolor}8 &
\cellcolor{highcolor}9 &
\cellcolor{highcolor}5 \\
& \cellcolor{highcolor}Mid & & & & & & \\
\hline

\cellcolor{lowcolor}\multirow{3}{*}{\textbf{T13}}
& \cellcolor{lowcolor}Low & & & & & & \\
& \cellcolor{lowcolor}High &
\cellcolor{lowcolor}73.33 &
\cellcolor{lowcolor}75 &
\cellcolor{lowcolor}39 &
\cellcolor{lowcolor}11 &
\cellcolor{lowcolor}25 &
\cellcolor{lowcolor}9 \\
& \cellcolor{lowcolor}Low & & & & & & \\
\hline

\cellcolor{midcolor}\multirow{3}{*}{\textbf{T14}}
& \cellcolor{midcolor}High & & & & & & \\
& \cellcolor{midcolor}Low &
\cellcolor{midcolor}81 &
\cellcolor{midcolor}32 &
\cellcolor{midcolor}9 &
\cellcolor{midcolor}6 &
\cellcolor{midcolor}17 &
\cellcolor{midcolor}3 \\
& \cellcolor{midcolor}Mid & & & & & & \\
\hline

\end{tabular}

\vspace{2mm}

\begin{tabular}{cl}
\cellcolor{lowcolor}\rule{0pt}{9pt}\hspace{12pt}
& Low-performing team \\[1pt]
\cellcolor{midcolor}\rule{0pt}{9pt}\hspace{12pt}
& Mid-performing team \\[1pt]
\cellcolor{highcolor}\rule{0pt}{9pt}\hspace{12pt}
& High-performing team
\end{tabular}

\end{table*}

Table \ref{tab:team_it} presents the distribution of IT connections across all 14 teams. Teams are color-coded according to their academic performance classification (high-, mid-, and low-performing). The figure shows variation among individual teams in both the overall number of IT connections and the distribution of connections across the three levels of integration.

Integrative thinking (IT) was observed across all teams; however, the frequency and depth of integration varied among teams. Some teams demonstrated a greater number of total IT connections and substantial-level integrations, whereas others demonstrated fewer connections or a greater proportion of superficial-level integrations. This variation indicates that students across all performance teams engaged in integrative thinking, although the extent and depth of integration differed across teams.
\begin{table}[ht]
\centering
\caption{Low Performing team Avegrage - IT Table}
\label{tab:it_connections2}
\begin{tabular}{lcccc}
\hline
\textbf{IT Connection} & \textbf{Superficial} & \textbf{Intermediate} & \textbf{Substantial} & \textbf{Total} \\
\hline
SBT-MBT & 2 & 1 & 5 & 8 \\
SBT-DBT & 2 & 2 & 5 & 9 \\
SBT-MER & 1 & 2 & 4 & 7 \\
SBT-CT  & 1 & 0 & 3 & 4 \\
MBT-DBT & 1 & 1 & 3 & 5 \\
MBT-MER & 3 & 1 & 1 & 5 \\
MBT-CT  & 1 & 1 & 3 & 5 \\
DBT-MER & 2 & 1 & 4 & 7 \\
DBT-CT  & 1 & 0 & 0 & 1 \\
MER-CT  & 1 & 0 & 0 & 1 \\
\hline
\textbf{Total} & \textbf{15} & \textbf{9} & \textbf{28} & \textbf{52} \\
\hline
\end{tabular}
\end{table}

\begin{table}[ht]
\centering
\caption{Mid Performing team Average - IT Table}
\label{tab:it_connections3}
\begin{tabular}{lcccc}
\hline
\textbf{IT Connection} & \textbf{Superficial} & \textbf{Intermediate} & \textbf{Substantial} & \textbf{Total} \\
\hline
SBT-MBT & 1 & 1 & 5 & 7 \\
SBT-DBT & 1 & 1 & 3 & 5 \\
SBT-MER & 1 & 2 & 2 & 5 \\
SBT-CT  & 0 & 0 & 2 & 2 \\
MBT-DBT & 1 & 1 & 2 & 4 \\
MBT-MER & 2 & 2 & 0 & 4 \\
MBT-CT  & 1 & 1 & 2 & 4 \\
DBT-MER & 2 & 1 & 1 & 4 \\
DBT-CT  & 0 & 0 & 1 & 1 \\
MER-CT  & 0 & 0 & 0 & 0 \\
\hline
\textbf{Total} & \textbf{9} & \textbf{9} & \textbf{18} & \textbf{36} \\
\hline
\end{tabular}
\end{table}

\begin{table}[ht]
\centering
\caption{High Performing team Average- IT Table}
\label{tab:it_connections4}
\begin{tabular}{lcccc}
\hline
\textbf{IT Connection} & \textbf{Superficial} & \textbf{Intermediate} & \textbf{Substantial} & \textbf{Total} \\
\hline
SBT-MBT & 2 & 2 & 7 & 11 \\
SBT-DBT & 2 & 2 & 4 & 8 \\
SBT-MER & 4 & 2 & 1 & 7 \\
SBT-CT  & 1 & 1 & 4 & 6 \\
MBT-DBT & 2 & 1 & 2 & 5 \\
MBT-MER & 4 & 1 & 1 & 6 \\
MBT-CT  & 1 & 2 & 5 & 8 \\
DBT-MER & 2 & 1 & 3 & 6 \\
DBT-CT  & 0 & 1 & 0 & 1 \\
MER-CT  & 1 & 1 & 0 & 2 \\
\hline
\textbf{Total} & \textbf{19} & \textbf{14} & \textbf{27} & \textbf{60} \\
\hline
\end{tabular}
\end{table}
To compare patterns across academic performance teams, team-level IT counts were averaged within each performance category. High-performing teams demonstrated the highest average number of IT connections (60), followed by low-performing teams (52) and mid-performing teams (36). Although these differences should be interpreted cautiously given the small number of teams, the pattern suggests that high-performing teams may have engaged in more frequent connections among different ways of thinking. One possible explanation is that high-performing teams may have more consistently integrated five ways of thinking in the WoT4EDP framework when developing and explaining their solutions, resulting in a greater number of identifiable IT connections. The relatively high number of connections among low-performing teams (Table~\ref{tab:it_connections2}), however, suggests that the relationship between academic performance and integrative thinking may not be linear.

Although high-performing teams demonstrated the greatest overall frequency of IT connections, the difference between high- and low-performing teams was relatively small (Table~\ref{tab:it_connections2} and Table~\ref{tab:it_connections4}). The similarity between these teams suggests that academic performance and the frequency of integrative thinking are not directly equivalent.  In particular, teams classified as low-performing based on final examination scores were still capable of demonstrating substantial integration among five ways of thinking in WOT4EDP framework within their design explanations.

When examining the depth of integration, both high- and low-performing teams demonstrated the highest average number of substantial (Level 3) IT connections, with an average of 26 connections. In comparison, mid-performing teams reported a lower average of 18 substantial IT connections (Table~\ref{tab:it_connections3}). Intermediate (Level 2) connections were relatively similar across teams, with low-, mid-, and high-performing teams demonstrating averages of 9, 10, and 12 connections. Similarly, superficial (Level 1) connections were observed across all teams, with low-, mid-, and high-performing teams demonstrating averages of 15, 11, and 17 connections.

Together, these results demonstrate that while high-performing teams showed the highest overall frequency of integrative thinking, substantial integration was not limited to academically higher-performing teams. The relationship between examperformance in the course and integrative thinking in the CPDE appears to be complex, suggesting that academic achievement alone may not fully capture students' ability to coordinate multiple ways of thinking within engineering design activities.

The distribution of the types of connections across the different ways of thinking revealed generally similar patterns among the three performance teams (See Table~\ref{tab:it_connections2} and Table~\ref{tab:it_connections4}). Across high-, mid-, and low-performing teams, students most frequently integrated Science-Based Thinking with Design-Based Thinking (SBT–DBT) and Science-Based Thinking with Math-Based Thinking (SBT–MBT). SBT–DBT connections occurred eight times among high-and low-performing teams, and five times among mid-performing teams. Similarly, SBT–MBT connections occurred 11 times among high-performing teams, seven times each among mid- and low-performing teams respectively. Connections between Design-Based Thinking and Metacognitive Reflection (DBT–MER) were also observed relatively frequently, occurring six times in high-performing teams, four times in mid-performing teams, and seven times in low-performing teams. Science-Based Thinking–Metacognitive Reflection (SBT–MER) connections followed a similar pattern, occurring seven times among high- and low-performing teams and five times among mid-performing teams respectively.

Overall, these patterns suggest that students across all performance teams primarily integrated scientific reasoning with design, mathematical, and reflective thinking. The frequent SBT–DBT and SBT–MBT connections indicate that students regularly drew on scientific principles to inform design decisions and support quantitative analysis. At the same time, the relatively frequent DBT–MER and SBT–MER connections suggest that students were also engaged in reflection while developing and evaluating their designs. These patterns are consistent with the demands of engineering design-based physics tasks, which require students to apply physics concepts, use quantitative reasoning, evaluate design choices, and reflect on their reasoning and decision-making processes in engineering design tasks, thereby attending to the design-science connections \cite{subramaniam2025design}. 

Across the three levels of performance, the distribution of connection levels varied depending on the pair of ways of thinking being integrated. For SBT–DBT, substantial connections were more frequent than intermediate or superficial connections across all three performance teams. Low-, mid-, and high-performing teams demonstrated on average four substantial design-science connections, compared with about one or two intermediate and superficial connections. A similar pattern was observed for SBT–MBT, where substantial connections were more frequent than intermediate or superficial connections across the performance teams. Low-, mid-, and high-performing students demonstrated on average about six substantial connections, compared with about 1-2 intermediate orsuperficial connections on average.

In contrast, the pattern for SBT–MER differed across performance teams. Low-performing students demonstrated 4 substantial, 2 intermediate, and 1 superficial connection, whereas mid-performing students demonstrated 2 connections at substantial and intermediate but only 1 superficial. Among high-performing students, superficial connections (4) were more frequent than substantial (1) or intermediate (2) connections. Thus, while low-performing students exhibited more substantial than superficial SBT–MER connections, high-performing students showed the opposite pattern.

The DBT–MER connections also varied across performance teams. Low-performing students demonstrated 4 substantial, 1 intermediate, and 2 superficial connections. Mid-performing students demonstrated 1 substantial, 1 intermediate, and 2 superficial connections, while high-performing students demonstrated 3 substantial, 1 intermediate, and 2 superficial connections. These findings indicate that metacognitive reflection does not consistently increase in sophistication with overall performance, unlike the more consistent patterns observed for SBT–DBT and SBT–MBT.

In contrast, connections involving Computational Thinking (CT) were less frequent across all performance teams. Design-Based Thinking–Computational Thinking (DBT–CT) connections occurred only once in each performance team. Similarly, Computational Thinking–Metacognitive Reflection (CT–MER) connections occurred twice among high-performing teams and once among both mid- and low-performing teams. Although these connections were less common than those involving scientific, mathematical, design, and reflective thinking, their presence indicates that students incorporated computational and reflective approaches into their design reasoning in some instances. For the SBT–CT connections, however students demonstratedmore substantial connections across all three performance teams rather than intermediate or superficial connections. All teams, regardless of performance level demonstrated 2-4 substantial connections, compared with 0-1 intermediate or superficcial connections. A similar pattern was observed for MBT–CT connections. All teams, regardless of performance level demonstrated 2-5 substantial connections, compared with 1-2 intermediate orsuperficial connections. Thus, for both SBT–CT and MBT–CT, substantial connections were considerably more frequent than intermediate or superficial connections across all performance levels.

For MBT–DBT, students demonstrated more substantial connections than intermediate or superficial connections across all levels of performance. All teams demonstrated 2-3 substantial connections,  compared with one intermediate connection and 1-2 superficial connections. In contrast, MBT–MER connections showed greater variation across performance teams. Low-performing students demonstrated 1 substantial, 1 intermediate, and 3 superficial connections; mid-performing students demonstrated no substantial connections, with 2 intermediate and 2 superficial connections; and high-performing students demonstrated 1 substantial, 1 intermediate, and 4 superficial connections. Thus, MBT–DBT connections were more frequently substantial, whereas MBT–MER connections were predominantly superficial or intermediate, with substantial connections occurring infrequently across all three performance teams.





\subsection{RQ2: Qualitative Analysis of Integrative Thinking Connections}
\begin{figure}
    \centering
    \includegraphics[width=1\linewidth]{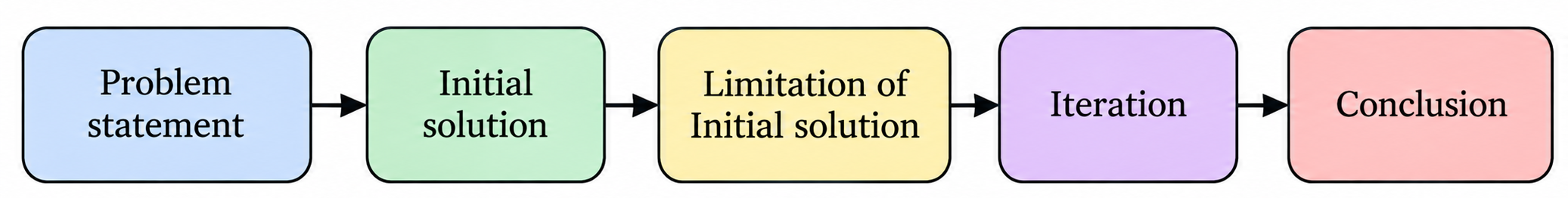}
    \caption{Stages of ED}
    \label{fig: ED stages}
\end{figure}

Our findings are organized to illustrate how each team’s integrative thinking develops across the stages of the ED process~\ref{fig: ED stages}. 
In each subsection we present a thick description of selected essay paragraphs, highlighting how different ways of thinking — science, mathematics, computation, design, and metacognition — are interconnected within students’ written work. Together, these analyses show how integrative thinking emerges and evolves across the artifact, rather than being confined to isolated parts of the solution process.
\subsubsection{\textbf{Problem Statement}}

We describe and analyze below the CPDE Problem-Statements constructed by three different teams: T01 (High-Performing Team), T09 (Mid-Performing Team), and T10 (Low-Perfoming Team). For each team we analyze the Integrative Thinking (IT) demonstrated their respective Problem Statements. Finally, we present a 'Cross-Team Comparison' of the Problem Statements of the three teams.

\subsubsubsection{High-Performing Team}

The problem statement composed by High-Performing Team, T01 is shown in Fig. ~\ref{HPTPS}. 
\begin{figure}[H]
\centering
\begin{tcolorbox}
\justify{`` According to the MLB, the chances of hitting a home run is approximately 12.10\%. Hitting a home run is not as easy as it seems. Using our physics knowledge that we have acquired from this class, we can calculate speed of pitch and how much spin a pitcher puts on a ball in order to predict the outcome of the ball's trajectory. ''}   
\end{tcolorbox}
\caption{Problem statement composed by High-Performing Team, T01.}\label{HPTPS}
\end{figure}

This excerpt demonstrates SBT-MBT at Level 2 (Intermediate) because the students connect their \textit{``physics knowledge''} to specific quantities they can \textit{``calculate,''} including the \textit{``speed of [the] pitch''} and \textit{``how much spin a pitcher puts on a ball,''} to predict the ball's trajectory. SBT-MER at Level 1 (Superficial) is evident in the general reflection that they use \textit{``physics knowledge that we have acquired from this class,''} without explaining how this prior knowledge informs their reasoning. MBT-MER at Level 1 is reflected in the vague statement that \textit{``we can calculate''} relevant quantities without reflecting on the mathematical reasoning or equations used. SBT-DBT at Level 1 appears in connecting physics knowledge to the problem of predicting \textit{``the outcome of the ball's trajectory,''} but without discussing criteria, constraints, or refinement. Similarly, MBT-DBT at Level 1 is demonstrated by using calculations to address the trajectory problem without specifying equations or how the calculations shape the solution. Finally, DBT-MER at Level 1 is reflected in the general consideration of the challenge of predicting the ball's trajectory, without deeper reflection on design decisions or limitations.

\subsubsubsection{Mid-Performing Team}
The problem statement composed by the Mid-Performing team, T09 is shown in Fig. ~\ref{MPTPS}. 
\begin{figure}[H]
\centering
\begin{tcolorbox}
\justify{`` With the world of toys becoming more and more competitive, companies like Duncan's Toy Chest are doing whatever they can to make better and better toys. Duncan's Toy Chest has commissioned
our firm to create a toy catapult that is better than any other before it. This catapult must be mostly plastic with foam pieces in order to minimize any possibility of harm to the user. This includes a soft projectile and a covered spring to initiate the launching process. While it would be simple to make a
trebuchet like toy that launches the projectile by using a mass on the end instead of the spring, Duncan's Toy Chest has requested that our team find the final distance of the bean bag from the
point of launch, after it is launched. The reasoning for the maximum height to be found, is so that the company can ensure parents that the toy is safe to use indoors and outdoors. ''}   
\end{tcolorbox}
\caption{Problem statement composed by Mid-Performing Team, T09.}\label{MPTPS}
\end{figure}

Team T09's problem statement clearly situates the ED-task within both a defined user context and relevant scientific principles. The students identify the client, \textit{Duncan's Toy Chest}, and articulate the purpose of the design in relation to solving a physics-based problem. They further incorporate scientific thinking by referencing projectile motion to determine the \textit{maximum height}, demonstrating how physics concepts inform the design process. The problem statement also reflects attention to design constraints, particularly related to safety. The students emphasize that "\textit{the company can ensure parents that the toy is safe to use indoors and outdoors}", indicating an awareness of real-world use conditions. Together, these elements show how the team supports its design decisions regarding user needs and safety constraints, with scientific thinking. This is consistent with Level 3 IT of SBT-DBT as per the criteria articulated in Table ~\ref{tab:IT_examples}.

\subsubsubsection{Low-Performing Team}

The problem statement composed by the Low-Performing team, T10 is shown in Fig.~\ref{LPTPS}. 
\begin{figure}[H]
\centering
\begin{tcolorbox}
\justify{`` Bulletproof vests can be traced all the way back to the 1500s. With advancements in technology,
bulletproof vests are now smaller and more effective than ever before, and guns have much more power than they did in the 1500s. This fact still remains, however: bulletproof vests stop bullets in
an extremely short amount of time and distance, and in this case, we have to find a force that will stop the bullet in 3.8 cm from a bulletproof vest, which is a constraint of the problem. The local
police force has tasked us with solving the following problem: How much force is required to stop a 0.0075 kg bullet shot from 2 m away from a vest, traveling at
1320 km/hr and taking 0.00073 seconds to travel the distance, in 3.8 cm after hitting the vest. ''}   
\end{tcolorbox}
\caption{Problem statement composed by Low-Performing Team, T10.}\label{LPTPS}
\end{figure}

The students in Team T10 demonstrate multiple forms of integrative thinking by framing an authentic engineering design problem. They explain that \textit{"bulletproof vests stop bullets in an extremely short amount of time and distance"} and connect this principle to the design requirement of determining \textit{"the force that will stop the bullet in 3.8 cm,"} showing a meaningful integration of SBT and DBT at Level 2, as physics is used to define the design constraints rather than justify a specific solution. They further integrate SBT and MBT by describing the physical situation using measurable quantities, including the bullet's mass, velocity, travel time, and stopping distance, demonstrating that scientific reasoning is supported by relevant mathematical information (Level 2). The strongest integration occurs between MBT and DBT (Level 3), where explicit numerical values \textit{"0.0075 kg," "1320 km/hr," "0.00073 seconds,"} and \textit{"3.8 cm"} are embedded directly into the engineering problem, establishing quantitative design criteria that define the solution space rather than merely providing supporting calculations.

\subsubsubsection{Cross-Team Comparison}

Across the three teams representing the three performance levels, we find that the SBT-DBT and SBT-MBT connections were present at varying levels of depth, but the teams differed in how explicitly these connections informed the engineering problem. The High-performing team demonstrated SBT-MBT at Level 2, while its connections involving design and reflection remained at Level 1. In contrast, the Mid-performing team demonstrated the deepest SBT-DBT connection at Level 3, linking scientific principles to user needs and safety constraints. The Low-performing team demonstrated Level 2 connections for both SBT-DBT and SBT-MBT and reached Level 3 for MBT-DBT by embedding quantitative criteria directly into the design problem. Thus, while the High-performing team showed intermediate integration primarily between science and mathematics, the Mid- and Low-performing teams demonstrated deeper integration involving design thinking, with the Low-performing team showing the strongest quantitative-design connection. Across teams, MER connections remained limited or were not evident beyond Level 1, indicating comparatively less developed integration of reflection with the other ways of thinking.

\subsubsection{\textbf{Initial Solution}}

We describe and analyze below the Initial Solution constructed by three teams. For each team we analyze the Integrative Thinking (IT) demonstrated their initial solutions. Finally, we present a 'Cross-Team Comparison' of the initial solutions of the three teams.

\subsubsubsection{High-Performing Team}

The Initial Solution composed by High-Performing Team, T01 is shown in Fig. ~\ref{HPTIS}. 

\begin{figure}[H]
\centering
\begin{tcolorbox}
\justify{`` For our first attempt, we just tried to figure out roughly what the calculations would look like in terms of calculating launch momentum after the ball is struck by the bat. We neglected to take into account the fact that the force is what’s being applied at an angle, and not that the launch momentum is traveling at the impact angle. This is something we didn’t realize through iteration one, but was improved in iteration two. ''}   
\end{tcolorbox}
\caption{Initial Solution composed by the High-Performing Team, T01.}\label{HPTIS}
\end{figure}

This excerpt demonstrates DBT-MER at Level 3 (Substantial) because the students explicitly reflect on why their initial approach was incorrect and how it was improved through iteration: \textit{``This is something we didn't realize through iteration one, but was improved in iteration two.''} MBT-DBT at Level 2 (Intermediate) is evident in their attempt to determine \textit{``what the calculations would look like''} for \textit{``calculating launch momentum''} and their use of this mathematical thinking to refine the solution across iterations. SBT-MBT at Level 2 is demonstrated through the connection between the physics concepts of \textit{``force''} and \textit{``launch momentum''} and the calculations used to model the impact. SBT-DBT at Level 3 is evident because the students identify a specific physics-based problem in their initial design-\textit{``the force is what's being applied at an angle, and not that the launch momentum is traveling at the impact angle''}-and use this understanding to improve the solution in the second iteration. SBT-MER at Level 1 is reflected in the acknowledgment that they \textit{``didn't realize''} the physics issue initially, without further reflection on how their physics understanding developed. Finally, MBT-MER at Level 1 appears in their reflection on the initial calculations without explaining the mathematical thinking, equations, or values underlying the correction.

\subsubsubsection{Mid-Performing Team}

The Initial solution composed by Mid-Performing Team, T09 is shown in Fig. ~\ref{MPTIS}. 

\begin{figure}[H]
\centering
\begin{tcolorbox}
\justify{`` Ever since catapults were first used in battle, young children throughout the lands would use toys modeled after them. This fascination with medieval siege weaponry has lasted the test of time, as catapult toys are extremely common to this day. The company Duncan's Toy Chest has requested the creation of a toy catapult better than any before it. This toy would be used by children of ages 5 and up, and by their parents. The size and mass of the projectile will be constant at 40 grams , and the force of gravity will be constant at 9.8 . The catapult arm will be 0.30 m with the pivot point at the center of the arm. For safety, a cover for the spring, and a pad on the arm of the catapult will
both be included. The projectile will be a soft bean bag to help ensure no harm will be caused on impact. The spring will have a relaxed length of 5 cm, which will constitute a final launching angle of
20°. This launch angle will determine the maximum height of the projectile after launch. ''}   
\end{tcolorbox}
\caption{Initial Solution composed by Mid-Performing Team, T09.}\label{MPTIS}
\end{figure}

Building on their initial framing,  Team T09 situates the problem within a clear client and user context, \textit{used by children of ages 5 and up,} while simultaneously defining key variables and assigning values aligned with the problem constraints (e.g., projectile mass of 40 grams and gravitational acceleration of 9.8 m/s$^2$). This reflects a purposeful alignment of mathematical representations with design requirements. The Level 3 Design-based Thinking - Science-based Thinking (DBT-SBT) integration is evident in their attention to safety as a design constraint, describing the projectile as a \textit{soft bean bag to ensure no harm,} while grounding this decision in the scientific context of projectile motion. Furthermore, a Level 3 Science-based Thinking - Math-based Thinking (SBT-MBT) connection is reflected in how they define variables based on scientific relationships. For example, by stating a \textit{launching angle of 20$^\circ$} and linking it to the \textit{maximum height of the projectile,} they demonstrate how mathematical parameters are derived from and used within scientific concepts to guide the design solution. Together, these connections illustrate a coherent and well-integrated approach.

\begin{figure}[hbt]
    \centering
    \includegraphics[width=1.0\linewidth]{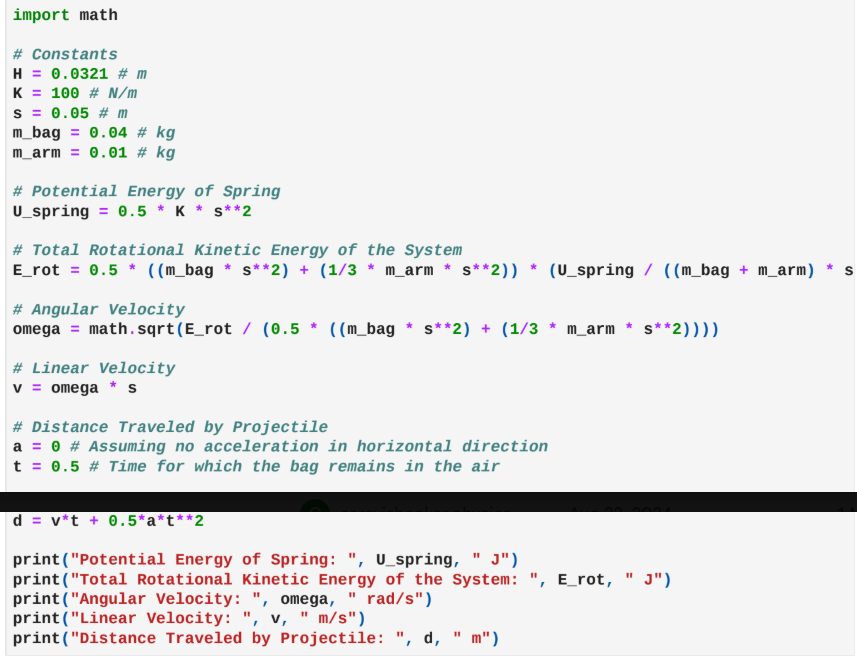}
    \caption{Student code by Mid-Performing Team, T09 in their Initial Solution.}
    \label{fig:Student code- Mid}
\end{figure}
The student code cell (Fig.~\ref{fig:Student code- Mid}) demonstrates substantial MBT-CT integration (Level 3) because the code implements mathematical equations to perform quantitative analysis. By translating mathematical relationships into code. The code also demonstrates substantial SBT-CT integration (Level 3), as the implemented equations are based on physics relationships and are used to model the physical behavior of the system. Additionally, the code demonstrates substantial DBT-CT integration (Level 3) because comments within the code describe design assumptions, showing that computational modeling was connected to decisions made during the design process.

\subsubsubsection{Low-Performing Team}

The Initial solution composed by Low-Performing Team, T10 is shown in Fig. ~\ref{LPTIS}. 

\begin{figure}[H]
\centering
\begin{tcolorbox}
\justify{`` From researching the statistics involving the bullet and bulletproof vests, we found that a 9mm bullet has a mass of around 0.0075 kg and that a normal distance for a bulletproof vest to stop a bullet in is 0.038 meters, and that the bullet would be traveling at around 1320 km/h. From using this data, we measured the time it would take for the bullet to stop once it initially came in contact with the bullet proof vest, which took about 2.07x10\^-4 seconds. This was solved through dividing the initial velocity by two to find the average velocity as the bullet comes to a stop, and dividing the distance by this average velocity. Then after using the momentum principle to solve for the force, we found that after assuming the force being applied from the vest is constant, that a force of 13267.785 N needs to be applied to stop this bullet. However, one limitation of this attempt at the solution is that it does not factor in other factors that would occur in a real life situation. For example, one thing that would happen in this situation would involve the bullet flattening as it impacts with the vest, which would change the force required from the vest. ''}   
\end{tcolorbox}
\caption{Initial Solution composed by Low-Performing Team, T10.}\label{LPTIS}
\end{figure}

The students in Team T10, demonstrate substantial integrative thinking across multiple connections by using scientific principles, mathematical analysis, design considerations, and reflection to develop and evaluate their solution. The integration of DBT and MER is demonstrated at Level 3 as students not only evaluate their solution but also identify limitations, stating that \textit{``one limitation of this attempt at the solution is that it does not factor in other factors that would occur in a real life situation,''} and explain how bullet deformation could affect the required force. They demonstrate MBT and DBT at Level 3 by using specific quantitative parameters, including \textit{``0.0075 kg,'' ``0.038 meters,''} and \textit{``1320 km/h,''} to define the design problem and calculate the required force for the bulletproof vest. The connection between DBT and SBT is also substantial (Level 3), as students apply physics principles, explaining that they used \textit{``the momentum principle to solve for the force''} and connect the scientific model to the design requirement of stopping the bullet. Additionally, the students demonstrate SBT-MER at Level 2 by reflecting on the assumptions of their physics model and identifying real-world factors that were not included. The integration of MBT-MER is also Level 2, as students describe how calculations were used to obtain the solution but provide limited reflection on how the mathematical model influences their understanding of the problem. Finally, the SBT-MBT connection reaches Level~3 because students explicitly connect the physics concept of momentum with mathematical calculations, using average velocity and stopping time to determine the force required to stop the bullet.

\subsubsubsection{Cross-Team Comparison}

Across the three teams, representing the three performance levels, integrative thinking in the initial solution becomes more developed as compared to Problem Statement, with all teams demonstrating connections among scientific, mathematical, design, and reflective thinking. The High-performing team demonstrates Level 2 SBT-MBT, Level 3 SBT-DBT, Level 1 SBT-MER, Level 2 MBT-DBT, Level 1 MBT-MER, and Level 3 DBT-MER. The Mid-performing team demonstrates Level 3 SBT-MBT and SBT-DBT, Level 3 MBT-DBT, Level 3 DBT-MER, and Level 3 SBT-CT, MBT-CT, and DBT-CT. The Low-performing team demonstrates Level 3 SBT-MBT and SBT-DBT, Level 3 MBT-DBT, Level 2 SBT-MER and MBT-MER, and Level 3 DBT-MER. Thus, all three teams demonstrate SBT-MBT, SBT-DBT, MBT-DBT, and DBT-MER, although the High-performing team shows lower levels for several of these connections. The Mid- and Low-performing teams demonstrate broader Level 3 integration, with the Mid-performing team additionally incorporating SBT-CT, MBT-CT, and DBT-CT at Level 3, while the Low-performing team demonstrates stronger integration involving reflection than the Mid-performing team.

\subsubsection{\textbf{Limitations of the Initial Solution }}

We describe and analyze below the limitations to the initial solution cited by three teams. For each team we analyze the Integrative Thinking (IT) demonstrated their descriptions of the limitations of their initial solutions. Finally, we present a 'Cross-Team Comparison' of the limitations of their initial solutions cited by the three teams.

\subsubsubsection{High-Performing Team}

We discuss what High-Performing team, T01 perceived to be an inherent limitation in their initial design solution. Fig. ~\ref{HPTLIS} shows the limitations cited by the High-Performing team.

\begin{figure}[H]
\centering
\begin{tcolorbox}
\justify{`` As stated above, the way that we handled the impact angle with respect to total momentum was originally incorrect, and this continued into iteration one. ''}   
\end{tcolorbox}
\caption{Limitations to Initial Solution cited by High-Performing Team, T01.}\label{HPTLIS}
\end{figure}

This excerpt demonstrates DBT-MER at Level 1 (Superficial) because the students briefly reflect that \textit{``the way that we handled the impact angle with respect to total momentum was originally incorrect''} and that the issue \textit{``continued into iteration one.''} This acknowledges a limitation in their initial solution, but the excerpt does not explain why the approach was incorrect, how it affected the solution, or how the problem was addressed. Therefore, the reflection remains superficial.

\subsubsubsection{Mid-Performing Team}

We discuss what Mid-Performing team, T09 perceived to be an inherent limitation in their initial design solution. Fig. ~\ref{MPTLIS} shows the limitations cited by the High-Performing team.

\begin{figure}[H]
\centering
\begin{tcolorbox}
\justify{`` Initially the original goal was to find the spring constant needed to get the bean bag to travel a target distance of 2 m. After many hours of working through the problem, constantly changing our processes and trying to figure out ways to solve the problem, we finally determined that with our current knowledge on the laws of physics, finding such a niche spring would not be practical. Adding that to the fact that in the real world, most springs have set stiffnesses as it is, we decided that it
would be more practical to determine the maximum height of the projectile after it launches, when the spring has a given constant. This change was chosen as it still allows us to make an indoor
friendly toy for children of all ages to enjoy. ''}   
\end{tcolorbox}
\caption{Limitations to Initial Solution cited by Mid-Performing Team, T09.}\label{MPTLIS}
\end{figure}

The team demonstrates a clear Level 3 DBT-SBT integration, as they use the \textit{laws of physics} to evaluate the feasibility of their initial design goal. Their reflection extends into Level 3 Science-based Thinking - Metacognitive Reflection (SBT-MER), as they describe how they were \textit{constantly changing our processes} in an attempt to determine a spring constant that would achieve a 2\,m distance, while recognizing the constraint that \textit{most springs have set stiffnesses}. This moment reflects a shift in their approach, as the recognition of a scientific constraint further demonstrates Level 3 Design-based Thinking - Metacognitive Reflection (DBT-MER) integration, with the team critically evaluating and revising their strategy. Rather than persisting with an impractical goal, they justify a strategic pivot by determining that it is \textit{more practical to determine the maximum height} for a given spring constant. In doing so, they align the physics of \textit{projectile launches} with the practical design objective of creating an \textit{indoor friendly toy}. This reasoning demonstrates a high level of integrative thinking, as the team uses constraints productively to guide decision-making while balancing physical feasibility with user-centered design considerations, such as safety for \textit{children of all ages.}

\subsubsubsection{Low-Performing Team}

We discuss what Low-Performing team, T10 perceived to be an inherent limitation in their initial design solution. Fig. ~\ref{LPTLIS} shows the limitations cited by the High-Performing team.

\begin{figure}[H]
\centering
\begin{tcolorbox}
\justify{`` A limitation of this iteration is the the bullet only compresses to half of its original length, which is
not entirely accurate to a real bullet. ''}   
\end{tcolorbox}
\caption{Limitations to Initial Solution cited by Low-Performing Team, T10.}\label{LPTLIS}
\end{figure}

This excerpt shows low-performing team, T10's awareness that their model did not fully represent a real world scenario. The team demonstrated superficial DBT-MER integration (Level 1) by identifying a limitation in their design model without explaining how this reflection influenced their design process. Students acknowledged that \textit{“the bullet only compresses to half of its original length.} However, they did not discuss how this limitation would affect their solution or lead to modifications in their design approach.

\begin{figure}[hbt]
    \centering
    \includegraphics[width= 1\linewidth]{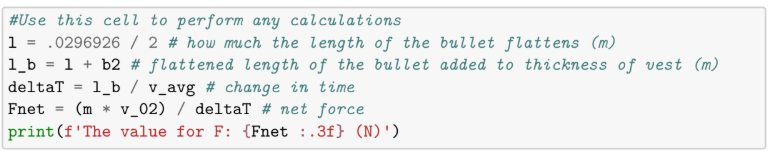}
    \caption{Student code by Low-Performing Team, T10 in their Initial Solution.}
    \label{fig:Student code-low}
\end{figure}

This student code cell (Fig. ~\ref{fig:Student code-low}) demonstrates substantial MBT-CT integration (Level 3) by using computational tools to implement mathematical equations for analysis. It also demonstrates substantial SBT-CT integration (Level 3), as the code applies physics relationships within a computational model to investigate the design problem.
\textit{"The following is the code for the graph. We utilized Matlab to create the graph, as our team is extremely familiar with the software."}

The team demonstrated superficial MBT-CT integration (Level 1) because they mentioned using MATLAB to create the graph but did not explain how computational tools supported their mathematical analysis. Students stated that \textit{“we utilized Matlab to create the graph, as our team is extremely familiar with the software,”} indicating that MATLAB was used mainly as a graphing tool rather than to develop mathematical understanding. The MER-CT and MER-MBT connections were also superficial (Level 1), as students reflected on their familiarity with the software but did not discuss how computational or mathematical reasoning informed their design decisions.

\subsubsubsection{Cross-Team Comparison}

Across the three teams, representing the three levels of performance, the depth of integrative thinking in identifying limitations varies substantially. The High-performing team demonstrates only Level 1 DBT-MER, as the students identify an error in their approach without explaining its consequences or how it was addressed. The Mid-performing team demonstrates the deepest integration, with Level 3 SBT-DBT, SBT-MER, and DBT-MER, as scientific constraints are used to evaluate the feasibility of the original design goal and guide a change in the design strategy. In contrast, the Low-performing team demonstrates Level 1 DBT-MER, MBT-CT, MER-CT, and MBT-MER, while SBT-CT is identified at Level 3 based on the computational implementation of physics relationships. Thus, the High- and Low-performing teams primarily demonstrate superficial reflection on limitations, whereas the Mid-performing team demonstrates substantially deeper integration by using scientific and design considerations to evaluate and revise its initial approach. The Low-performing team additionally incorporates computational thinking, but these connections are primarily related to implementation and software use rather than reflection on how the computational or mathematical reasoning influenced the design.

\subsubsection{\textbf{Iteration}}

The teams were required to iterate their initial design solutions. We describe and analyze below ways in which each team iterated upon its initial solution. For each team we analyze the Integrative Thinking (IT) demonstrated in the iteration of their initial solutions. Finally, we present a 'Cross-Team Comparison' of the iterations of the initial solutions cited by the three teams.

\subsubsubsection{High-Performing Team}

We discuss how students in the High-Performing team, T01 iterated on their initial solution. Fig. ~\ref{HPTIT} shows how the High-Performing team described the reason for iterating its initial solution.

\begin{figure}[H]
\centering
\begin{tcolorbox}
\justify{`` In this iteration, we introduced an entire solution to the first part we we’re attempting to look at, which is how angle of impact effects the trajectory of the ball in the field. As stated in our first attempt of the solution, our total momentum calculations were wrong, and so we got an answer that somewhat made sense, but wasn't totally correct. ''}   
\end{tcolorbox}
\caption{Reason for Iteration provided by High-Performing Team, T01.}\label{HPTIT}
\end{figure}

This excerpt demonstrates SBT-DBT at Level 1 (Superficial) because the students identify the physics-related problem of \textit{``how angle of impact affects the trajectory of the ball''} while describing their attempt to develop \textit{``an entire solution,''} but they do not explain the underlying physics or discuss criteria, constraints, or refinement. SBT-MER at Level 1 is evident in their acknowledgment that their \textit{``total momentum calculations were wrong''} and that the result \textit{``somewhat made sense, but wasn't totally correct,''} without explaining how their physics understanding changed or contributed to solving the problem. Similarly, DBT-MER at Level 1 is demonstrated by reflecting that the initial solution was \textit{``not totally correct,''} but without explaining why the solution failed, what design limitations were identified, or how the solution was subsequently improved.

Students in the High-Performance Team continued to describe their iteration of the initial solution in Fig. ~\ref{HPTIT2}.

\begin{figure}[H]
\centering
\begin{tcolorbox}
\justify{`` As for our second part, trying to see how the perpendicular distance (which we referred to location on the ball for simplicity) the bat struck the ball from impacted launch distances was a bit more of a struggle, but we did it this way. In our code we find the rotational speed from the net angular momentum with the given perpendicular distance. From this we can then use energy principle we 6 also were able to find rotational energy, and therefor translational energy. From the translational kinetic energy, we can get a launch velocity and then momentum to simulate the trajectory following the impact. ''}   
\end{tcolorbox}
\caption{Continued description of the Iteration provided by High-Performing Team, T01.}\label{HPTIT2}
\end{figure}

This excerpt demonstrates SBT–CT at Level 2 (Intermediate) because the students explicitly connect physics concepts to computational use, stating, textit{“In our code we find the rotational speed from the net angular momentum”} and then use the \textit{“energy principle”} to determine rotational energy, translational energy, launch velocity, and momentum. SBT–DBT at Level 2 is evident in their explanation of how the physics-based model is used to investigate how \textit{“perpendicular distance”} affects \textit{“launch distances.”} SBT–MER at Level 2 appears as they reflect on the challenge—\textit{“was a bit more of a struggle”}—and explain the physics-based approach they used to address it. CT–DBT at Level 2 is demonstrated by \textit{“In our code”} and their explanation of how the computational process contributes to simulating the trajectory. CT–MER at Level 2 is evident because they explain how code supports their reasoning through a sequence of calculations from rotational speed to launch velocity and momentum. Finally, DBT–MER at Level 1 is reflected in the brief statement that the task \textit{“was a bit more of a struggle,”} which acknowledges difficulty but does not explain design decisions, limitations, or why the chosen approach works.

\subsubsubsection{Mid Performing Team}

We discuss how Mid-Performing team, T09 iterated on its initial solution. Fig. ~\ref{MPTIT} shows how the High-Performing team described the reason for iterating its initial solution.

\begin{figure}[H]
\centering
\begin{tcolorbox}
\justify{`` \textit{Client}: Duncan's Toy Chest.
\textit{User}: Parents and children.
Assumption: The size and mass of the projectile will be constant, and so will the force of gravity. \\
\textit{Constants}: Mass of projectile = 40 g, Catapult arm = 0.30 m, Spring Constant = 100 N/m. \\
\textit{Criteria}: Toy must be safe enough for kids 5 years old and up to be able to use. This includes a cover for the spring, and a pad on the arm of the catapult. It also means a light, soft projectile. \\
\textit{Constraints}: Design must include the use of a spring as the main method of launching the projectile. Projectile mass will be 40 g in order to allow for a "fun" distance for the kids, but still light enough to impact with low force. Total size of catapult must be small enough to transport, so the max length of the arm (the largest portion of the catapult) cannot be longer than 0.3 m. \\
\textit{Iterative process}: Throughout the process we determined that the optimal way to work on this problem will be to find the distance of the bean bag when the acted on by known constants, instead of trying to find the constant, k for the spring force. This is due to the extreme complexity of this process. \\
\textit{Trade-offs}: The light-weight material of the toy (mostly plastics) can cause the toy to be fragile
unless made from more expensive types of plastic. ''}   
\end{tcolorbox}
\caption{Description of the Iteration provided by Mid-Performing Team, T09.}\label{MPTIT}
\end{figure}

In their solution iteration, the team demonstrated a Level 3 SBT-MBT connection, as they discussed the physics concept of \textit{spring force and projectile motion} while explicitly stating mathematical values such as a \textit{spring constant of 100 N/m} and a \textit{projectile mass of 40 g}. These discussions also support a Level 3 MBT-DBT connection, as the team explains how the \textit{numerical values of the given variables} relate to the final product’s \textit{low-force impact}. The SBT-DBT connection is also evident at Level 3, as the team presents a clear design rationale describing the \textit{use of a spring as the primary launching mechanism} to meet the safety requirements of the client, \textit{Duncan's Toy Chest}. In terms of reflection, the DBT-MER connection was interpreted as Level 3 because the team explains why the chosen solution works and identifies conditions under which it may not function as intended, particularly noting how \textit{safety features may partially inhibit movement}. However, reflections on scientific and mathematical reasoning are less developed. The SBT-MER connection was interpreted as Level 1, as the team identifies concepts such as \textit{gravity and mass} but provides only limited reflection. Similarly, the MBT-MER connection was also interpreted as Level 1, as terms such as \textit{constant K} are mentioned with minimal elaboration, indicating limited engagement with underlying scientific or mathematical reasoning.

\subsubsubsection{Low-Performing Team}

Students in the Low-Performing Team, T10 described their iteration of the initial solution in Fig. ~\ref{LPTIT}.

\begin{figure}[H]
\centering
\begin{tcolorbox}
\justify{`` For this iteration, we added more complexity to the first iteration by assuming that the bullet would flatten entirely after thinking more about how a bullet would realistically collide with a bulletproof vest in a real life scenario, since a bullet should flatten more than only half of the length. For how far the bullet traveled in this solution, it will have still traveled the same distance of 3.8 cm through the vest but also the full bullet would flatten, technically increasing the travel distance of the point mass of the bullet, leading to the result shown above. However, one limitation of this method is that most likely in the real world, a bullet might not flatten its entire length and could
be a smaller value, which would mean this solution could not fully be accurate in the real world. ''}   
\end{tcolorbox}
\caption{Description of the Iteration provided by Low-Performing Team, T10.}\label{LPTIT}
\end{figure}

In their solution iteration, the team demonstrated a Level 3 DBT-MER connection, as they reflected on how their design assumption changed from the previous iteration and explained why they decided to model the bullet as flattening entirely. Their reasoning was grounded in the realism of the design scenario, as they reconsidered how a bullet would \textit{“realistically collide with a bulletproof vest”} and recognized that the original assumption of partial flattening did not adequately represent the intended real-world situation. The team also identified a limitation of this design decision, noting that a real bullet may not flatten along its entire length, which could reduce the accuracy of their solution. The SBT-MER connection was interpreted as Level 3 because the team reflects on the physical behavior of the bullet during the collision and evaluates whether their assumption is realistic. Rather than simply identifying the physical phenomenon of bullet deformation, they explain how the assumed deformation affects their model and acknowledge that the actual behavior of a bullet may differ. The MBT-MER connection was also interpreted as Level 3, as the team explicitly reflects on how their physical assumption changes the mathematical representation of the problem. In particular, they retain the \textit{3.8 cm} distance traveled through the vest while explaining that complete flattening would technically increase the travel distance of the bullet's point mass. This demonstrates reflection on how a modeling assumption affects the mathematical solution and its accuracy.

The team also demonstrated Level 3 SBT-MBT and SBT-DBT connections. The SBT-MBT connection is evident because the physical behavior of the bullet is directly connected to the mathematical quantity used in the solution: the team explains that assuming complete flattening changes the effective travel distance of the point mass. The SBT-DBT connection is demonstrated through the team's use of a physical consideration to refine their solution design. Their decision to assume complete flattening was made specifically because they believed it would provide a more realistic representation of the bullet's collision with the vest. The MBT-DBT connection was interpreted as Level 3 because the team uses the mathematical consequence of the revised assumption to refine the solution, explaining how the increased effective travel distance is incorporated into the iteration. They also connect this mathematical model to a design limitation by recognizing that uncertainty in the actual amount of bullet deformation means the resulting solution may not fully represent a real-world scenario.

\subsubsubsection{Cross-Team Comparison}

Across the three teams, representing the three performance levels, the ways in which each team iterated their initial solution demonstrated more developed integrative thinking, although the depth and breadth of connections differ. The High-performing team demonstrates Level 2 SBT-MBT, SBT-DBT, and SBT-MER, with Level 1 DBT-MER, and also demonstrates Level 2 SBT-CT, DBT-CT, and MER-CT connections through the use of computational modeling. The Mid-performing team demonstrates Level 3 SBT-MBT and SBT-DBT, Level 1 SBT-MER, Level 3 MBT-DBT, Level 1 MBT-MER, and Level 3 DBT-MER. The Low-performing team demonstrates Level 3 SBT-MBT, SBT-DBT, and SBT-MER, as well as Level 3 MBT-DBT, MBT-MER, and DBT-MER. Thus, all three teams demonstrate SBT-MBT and SBT-DBT, but the High-performing team demonstrates these connections at Level 2, whereas the Mid- and Low-performing teams demonstrate them at Level 3. The Low-performing team shows the deepest integration involving reflection, with SBT-MER, MBT-MER, and DBT-MER all at Level 3, while the High- and Mid-performing teams show comparatively less developed reflective connections. The High-performing team is also the only team demonstrating CT-related connections in this iteration, although these remain at Level 2.

\subsubsection{\textbf{'Conclusion' of the CPDE Report}}
\subsubsubsection{High-Performing Team}

Students in the High-Performing Team, T01 described the Conclusion of their final report as shown in Fig. ~\ref{HPTC}.

\begin{figure}[H]
\centering
\begin{tcolorbox}
\justify{`` For the second part, its obvious that hitting the ball dead center, as to give it no spin, would maximize the distance traveled, but how much is it’s distance impacted by being hit a bit off from the center. From our graph, you can see that if you hit a bit past half of the radius (the radius of the ball is .0365 meters), it will result in a hit that doesn’t even reach the pitcher. However, if you hit just at the half radius mark, you’ll still send the ball 40 meters or more. This assures batters that as long as they hit reasonably close to dead center to minimize spinning the ball, that they still have a chance of hitting it decently far. ''}   
\end{tcolorbox}
\caption{Conclusion provided by the High-Performing Team, T10.}\label{HPTC}
\end{figure}

The SBT-MBT connection is at Level 2 because the response includes specific variables, shown by \textit{``the radius of the ball is .0365 meters,''} but it does not reach Level 3 since no formal physics principle is used to explain impact location, spin, and distance. The SBT-DBT connection is also at Level 2 because the statement \textit{``hitting the ball dead center... would maximize the distance traveled''} explains the idea conceptually, yet it does not present a clear design decision, constraint, or science concept in depth. For SBT-MER, the connection remains at Level 1 because the reflection is limited to reassurance, as seen in \textit{``This assures batters that as long as they hit reasonably close to dead center,''} and it does not explicitly reference formal physics principles such as the Angular Momentum Principle. The SBT-CT connection is at Level 2 since \textit{``From our graph, you can see''} indicates that code or computational tools were used, but the explanation does not include the logic, variables, or structure of the code itself. 

The MBT-DBT connection reaches Level 3 because mathematical results are used to refine the outcome, illustrated by \textit{``if you hit just at the half radius mark, you'll still send the ball 40 meters or more,''} which explains the constraints explicitly. The MBT-MER connection is at Level 1 because, although a numerical result is stated, such as \textit{``send the ball 40 meters or more,''} there is no explanation of why this mathematical approach was chosen or reflection on the methodology used. The MBT-CT connection is at Level 2 because data derived from computation are referenced in \textit{``From our graph... half radius mark,''} yet there are no direct references to the code that generated the data. 

The DBT-MER connection achieves Level 3 because the response clearly identifies failure conditions, shown by \textit{``if you hit a bit past half of the radius... it will result in a hit that doesn't even reach the pitcher,''} which explains when and why the design no longer works. The DBT-CT connection is at Level 2 because \textit{``From our graph, you can see''} suggests computational involvement in evaluating the design, but no written code or logical structure is provided. Finally, the MER-CT connection is at Level 2 because the explanation shows how computational output supports reasoning, as in \textit{``From our graph... This assures batters,''} but it does not describe how coding influenced the scientific thinking or decision-making process.

\subsubsubsection{Mid-Performing Team}
Students in the Mid-Performing Team, T09 described the Conclusion of their final report as shown in Fig. ~\ref{MPTC}.

\begin{figure}[H]
\centering
\begin{tcolorbox}
\justify{`` In the final solution to our problem, we were able to determine that due to the size of the toy, it will be able to be used indoors in a very safe manner. This allows our employer to deliver its promise to its customers of a safe and fun toy for children of all ages. There were some bumps along the road, and the intial trials did not work out as we expected. But, we were still able to deliver on the requested goal of a safe and fun toy that can be used indoors. ''}  
\end{tcolorbox}
\caption{Conclusion provided by the Mid-Performing Team, T09.}\label{MPTC}
\end{figure}

This team’s final solution demonstrates Level 3 DBT-MER IT, as it moves beyond a simple summary to provide a reflective account of the design’s success in relation to its initial constraints. The team explains why their solution is effective by explicitly linking the physical \textit{size of the toy} to the requirement of being \textit{used indoors}. They also describe the iterative nature of their process, acknowledging that \textit{initial trials did not work out as expected}, while emphasizing that they ultimately \textit{delivered on the requested goal of a safe and fun toy}. This reflection demonstrates clear reasoning about why the solution works and meets the criteria for the highest level of integrative thinking in this category.

\subsubsubsection{Low-Performing Team}
Students in the Low-Performing Team, T10 described the Conclusion of their final report as shown in Fig. ~\ref{LPTC}.

\begin{figure}[H]
\centering
\begin{tcolorbox}
\justify{`` Our final solution for the problem for the solution found overall how much force would need to be applied by a bulletproof vest to stop a bullet. When setting up the problem, we had to think of trade-offs that would make an adequate piece of body armor with using the bullet as the system and the armor as the surroundings, because the armor is what is applying the force on the bullet.
Furthermore for the measurements, we used a point particle system for the bullet since it would allow us to best find the constant external force on the bullet. If the force was smaller but the bullet
could penetrate more of the armor, the armor would be too bulky and unusable, so we had to find a more effective size that would be usable for the police force or other interested parties, leading us to finding this solution that we believe addresses all aspects of this problem.. ''}  
\end{tcolorbox}
\caption{Conclusion provided by the Low-Performing Team, T10.}\label{LPTC}
\end{figure}

In their final solution, this team demonstrated a Level 3 SBT-DBT connection, as they connected the physics concept of \textit{force} to the design requirements of a bulletproof vest. They explicitly considered the armor as the surroundings and the bullet as the system, recognizing that the armor must apply an external force to the bullet in order to stop it. This physics-based consideration was incorporated into their design reasoning as they evaluated the trade-off between the force required to stop the bullet and the amount of armor needed. The team also demonstrated a Level 1 SBT-MBT connection because they identify the physics concept of \textit{constant external force} and mention the use of a point particle system, but they do not provide specific mathematical values, equations, or detailed calculations to explain how the mathematical model was used. Similarly, the MBT-DBT connection was interpreted as Level 1 because the team refers to using a point particle system to find the constant external force, but provides limited mathematical detail about how this quantity informed the final design.

The DBT-MER connection was interpreted as Level 3 because the team reflects in detail on the practicality of their solution and the trade-offs involved in designing usable body armor. They recognize that reducing the force would allow the bullet to penetrate farther, while increasing the amount of armor could make the vest too bulky and unusable. The team therefore explains why they sought a more effective size and connects this decision to the needs of the police force and other potential users. This demonstrates reflection on the criteria and practical limitations that influenced their design decisions. The SBT-MER connection was also interpreted as Level 3 because the team reflects on how their understanding of the force acting on the bullet informed their final solution. Their distinction between the bullet as the system and the armor as the surroundings demonstrates reflection on how the physical interaction operates within their model. They further evaluate the physical solution in relation to the practical goal of stopping the bullet while maintaining a usable size, showing how their physics reasoning contributed to their overall solution.

\subsubsubsection{Cross-Team Comparison}

Across the teams at the three performance levels, the conclusions demonstrate differences in both the breadth and depth of integrative thinking. The High-performing team demonstrates Level 2 SBT-MBT, SBT-DBT, SBT-CT, Level 1 SBT-MER, Level 3 MBT-DBT, Level 1 MBT-MER, Level 2 MBT-CT, Level 3 DBT-MER, Level 2 DBT-CT, and Level 2 MER-CT. The Mid-performing team demonstrates Level 3 DBT-MER, while no additional connections are identified in the conclusion. The Low-performing team demonstrates Level 1 SBT-MBT, Level 3 SBT-DBT, Level 3 SBT-MER, Level 1 MBT-DBT, and Level 3 DBT-MER. Thus, all three teams demonstrate DBT-MER at Level 3, indicating that reflection on design outcomes and practical considerations is a common feature of their conclusions. However, the High-performing team demonstrates the broadest range of connections, including multiple mathematical, computational, and reflective connections, whereas the Mid-performing team shows the narrowest range of integrative connections. The Low-performing team demonstrates fewer connections than the High-performing team but reaches Level 3 in SBT-DBT, SBT-MER, and DBT-MER, indicating deeper integration of scientific reasoning, design considerations, and reflection.

\section{DISCUSSION}

The quantitative analysis revealed that integrative thinking (IT) was evident across all three academic performance teams, although the frequency and depth of integration varied. High-performing teams demonstrated the highest average number of IT connections (60), followed by low-performing teams (52) and mid-performing teams (36). However, the relatively small difference between the high- and low-performing teams suggests that the frequency of integrative thinking does not correspond directly to academic performance. This pattern becomes more evident when considering the depth of integration. High- and low-performing teams demonstrated the same average number of substantial (Level 3) IT connections (27), whereas mid-performing teams demonstrated fewer substantial connections (18). Thus, substantial integrative thinking was not limited to academically higher-performing teams. These findings suggest that conventional measures of academic performance may not fully capture students’ ability to coordinate multiple ways of thinking during engineering design activities. This aligns with prior research in physics education research showing that quantitative problem-solving performance does not necessarily reflect students’ conceptual understanding \cite{morphew2018exploring,fakcharoenphol2015judgments}.

One possible explanation is that integrative thinking in CPDEs involves forms of thinking that differ from those emphasized in traditional assessments. Final examinations primarily assess individual conceptual understanding and problem-solving, whereas CPDEs require students to coordinate scientific, mathematical, computational, design, and metacognitive thinking while developing and evaluating an open-ended solution. Consequently, students who perform differently on conventional assessments may nevertheless demonstrate similar capacities to integrate multiple ways of thinking in an engineering design context. The relatively high number of substantial connections observed among low-performing teams provides particularly important evidence for this interpretation. Rather than indicating that lower academic performance is associated with weaker integrative thinking, the findings suggest that students’ ability to integrate different forms of thinking may not be adequately represented by examination performance alone.

At the same time, these results represent numerical differences in IT connections and should not be interpreted as evidence of statistically significant differences among academic performance teams. High-performing teams demonstrated the greatest overall number of IT connections, indicating that academic performance may be associated with the frequency with which students articulate integrations in their written work. However, because low-performing teams demonstrated a comparable number of total connections and the same average number of substantial connections as high-performing teams, the relationship appears more complex than a simple linear association. In particular, the mid-performing team demonstrated fewer total and substantial connections than both the high- and low-performing teams. This non-monotonic pattern suggests that factors beyond academic performance may shape how students integrate different ways of thinking during engineering design.

An important consideration is that the analysis was conducted using team-authored CPDE reports. Therefore, the identified IT connections represent the thinking documented by the team rather than the thinking of individual students. Collaborative processes may influence both the frequency and depth of the connections that become visible in the final artifact. For example, a particular student may contribute disproportionately to connecting scientific thinking with mathematical analysis or design decisions, while other members may contribute in different ways. Similarly, teams with comparable examination performance may differ in how they distribute responsibility for thinking, discussion, and writing. Consequently, the relationship between individual academic achievement and team-level integrative thinking cannot be established directly from the present data.

A second consideration concerns differences in the CPDE reports themselves. Although all teams completed the same assignment, the reports varied in length, level of detail, and style of explanation. Longer or more detailed reports provide more opportunities for students to make their thinking explicit and therefore may yield more observable IT connections. Conversely, a shorter report may contain fewer coded connections even when students engaged in comparable thinking during the design process. Thus, the frequency of IT connections should be interpreted as a measure of students’ documented integrative thinking, rather than a complete measure of the integrative thinking that occurred during the engineering design process.

The distribution of connections among different ways of thinking provides further insight into how students integrated their thinking. Across all three performance teams, SBT–DBT and SBT–MBT were among the most frequently observed connections. This pattern is consistent with the structure of engineering design-based physics activities, in which students must use scientific thinking to make and justify design decisions and employ mathematical thinking to analyze those decisions. Importantly, these connections were also frequently substantial. For SBT–DBT and SBT–MBT, substantial connections occurred more often than intermediate or superficial connections across the performance teams. This suggests that when students connected scientific thinking with design or mathematics, they often did more than simply reference the two ways of thinking together; they coordinated them to explain, justify, or evaluate their thinking.

The relatively frequent SBT–MER and DBT–MER connections further indicate that metacognitive reflection was not absent from students’ integrative thinking. Rather, students sometimes connected reflection with scientific thinking and design thinking as they evaluated their decisions and solutions. However, the depth of these connections varied across performance teams. For example, low-performing teams demonstrated more substantial than superficial SBT–MER connections, whereas high-performing teams demonstrated more superficial than substantial SBT–MER connections. A similar variation was observed for DBT–MER and MBT–MER connections. These patterns suggest that the sophistication of metacognitive integration does not necessarily increase with overall academic performance. Instead, students may engage in metacognitive thinking in qualitatively different ways depending on how reflection is connected with other ways of thinking.

In contrast, connections involving Computational Thinking (CT) were consistently less frequent. DBT–CT connections occurred only once in each performance team, while CT–MER connections occurred twice among high-performing teams and once among both mid- and low-performing teams. These findings suggest that computational thinking was less frequently integrated with other ways of thinking in students’ written explanations. However, the pattern for SBT–CT and MBT–CT provides an important qualification. Although these connections were relatively infrequent overall, when they occurred, they were predominantly substantial. For example, SBT–CT and MBT–CT connections showed substantially more Level 3 integrations than Level 1 or Level 2 integrations across the performance teams. This suggests that computational thinking may not have been routinely integrated throughout the CPDEs, but when students did explicitly connect computational thinking with scientific or mathematical thinking, those connections were often meaningful and substantial.

This distinction has important instructional implications. The findings do not simply suggest that students need to use computation more frequently. Rather, they suggest that instructional design could provide students with more opportunities to explicitly connect computational work to scientific thinking, mathematical thinking, and design decisions. Similarly, because metacognitive connections varied considerably in depth, students may benefit from prompts that make the relationship between reflection and other ways of thinking more explicit. For example, students could be encouraged to explain how a computational result changes a design decision, how a mathematical analysis supports a scientific interpretation, or how reflection on an unsuccessful design informs subsequent revisions. Such scaffolds may help make integrations that are otherwise implicit more visible in students’ thinking.

The qualitative case analyses further illuminate these quantitative patterns by showing that the depth of integrative thinking is characterized not simply by the presence of multiple ways of thinking, but by how students coordinate those ways of thinking within an explanation. Across the representative high-, mid-, and low-performing teams, students integrated scientific thinking with mathematical thinking, computational thinking, design thinking, and metacognitive thinking in different ways. The qualitative analysis therefore complements the quantitative findings by demonstrating that teams can produce connections involving the same pair of ways of thinking while differing in how explicitly one way of thinking is used to explain, justify, or evaluate another. Thus, examining only the frequency of connections would not fully capture the nature of students’ integrative thinking.

Taken together, the findings suggest that integrative thinking in engineering design-based physics is multidimensional. The frequency and depth of IT connections did not vary in the same way across academic performance teams. Although high-performing teams demonstrated the greatest overall frequency of integration, low-performing teams demonstrated a comparable number of substantial connections. Moreover, the distribution of connections across ways of thinking showed that some integrations, particularly SBT–DBT and SBT–MBT, were both frequent and often substantial, whereas computational integrations were less frequent but frequently substantial when they occurred. Metacognitive integrations were also evident across performance teams, although their depth varied.

These findings contribute to the operationalization of integrative thinking in engineering design-based physics by distinguishing both the frequency and depth of connections among ways of thinking. The WoT4EDP framework ~\cite{subramaniam2025stem} and the three-level rubric provide a systematic approach for examining not only whether students draw on multiple forms of thinking, but also how meaningfully they coordinate those forms of thinking within their design explanations. Overall, our findings demonstrate that engineering design experiences provide opportunities for students across different levels of academic performance to demonstrate integrative thinking and that such thinking cannot be fully characterized by academic performance alone.

\section{LIMITATIONS}

This study has several limitations that should be considered when interpreting the findings. First, the study was conducted within a single course context and institution and focused on one laboratory section, which may limit the transferability of the findings to other instructional settings. Future studies across multiple institutions, courses, and laboratory contexts are needed to examine the broader applicability of the WoT4EDP framework for characterizing integrative thinking.

Second, this study relied primarily on students’ written Computational Physics Design Essay (CPDE) artifacts to examine their demonstration of integrative thinking. Although these artifacts provide a valuable representation of students’ design processes and reasoning, they may not fully capture the cognitive processes students engaged in during the activity. Because the CPDE is a compilation document, students may have intentionally emphasized certain ways of thinking while minimizing others based on what they considered most relevant to communicate in the final report. As a result, some forms of thinking may be underrepresented in the written artifact even if they occurred during the design process. However, analyzing the final document remains necessary because it represents students’ formal articulation and integration of their design decisions, scientific reasoning, computational approaches, and reflections.

Third, the variability in CPDE report length and team collaboration may have influenced the number and visibility of identified integrative thinking connections. Differences in how teams documented their design processes, distributed writing responsibilities, and communicated their thinking could affect the extent to which integration was observable in the final artifacts.

Finally, the categorization of teams into high-, mid-, and low-performing teams was based on course performance. While this teaming provides a useful lens for examining patterns of integrative thinking, course performance may not capture all dimensions of students’ abilities, including creativity, collaboration, problem-solving approaches, or conceptual understanding that may contribute to engineering design performance.

\section{FUTURE WORK}

Future work will extend this study in several directions. First, the operationalization of Integrative Thinking (IT) within the Ways of Thinking for Engineering Design Processes (WoT4EDP) framework will be examined across a broader range of student teams and instructional contexts. In particular, future iterations of the course will incorporate modified instructional supports that explicitly emphasize metacognitive reflection and computational thinking (CT) to investigate whether targeted scaffolding influences the frequency, nature, and depth of students’ integrative connections.

Second, we plan to investigate the potential of large language models (LLMs) as an assistive tool for analyzing student design essays. Specifically, LLMs will be evaluated on their ability to identify and classify students’ ways of thinking according to the WoT4EDP framework and compared against expert human coding. This work will examine the extent to which LLM-based analysis can reliably capture the complex interactions among science-based, design-based, mathematics-based, computational, and metacognitive ways of thinking.

Third, future research will explore whether LLMs can support the application of the analytic rubric developed in this study for evaluating the depth of integrative thinking. By comparing LLM-generated evaluations with researcher coding, we aim to determine whether LLMs can provide consistent and meaningful assessments of superficial, intermediate, and substantial levels of integration while maintaining the interpretive depth required for qualitative analysis.

Finally, future work will extend the analysis of IT beyond pairwise connections among ways of thinking to examine broader combinations within the WoT4EDP framework. Because the framework includes five ways of thinking, students may demonstrate combinations involving two, three, four, or all five ways of thinking within a single meaningful unit of their design work. Examining these combinations across different sections of Computational Physics Design Essays could provide a more comprehensive understanding of how multiple forms of reasoning coexist and interact as students develop and communicate engineering design solutions. This direction would complement the analysis of connection depth presented in the current study by examining the broader patterns and structures through which students integrate multiple ways of thinking.

\section{CONCLUSIONS}

This study examined integrative thinking in students’ Computational Physics Design Essays (CPDEs) using the Ways of Thinking for Engineering Design Processes (WoT4EDP) framework. 

To address RQ1, How can the depth of integrative thinking in engineering design-based physics tasks be characterized and operationalized?, we developed a three-level analytic rubric to characterize the depth of connections among the five ways of thinking in the WoT4EDP framework: Science-Based Thinking (SBT), Design-Based Thinking (DBT), Math-Based Thinking (MBT), Computational Thinking (CT), and Metacognitive Reflection (MER). The rubric distinguishes superficial, intermediate, and substantial levels of integration based on the extent to which students connect and coordinate multiple ways of thinking within their written explanations. This operationalization makes the depth of integrative thinking observable and provides a systematic approach for identifying and characterizing differences in how students integrate ways of thinking within engineering design-based physics tasks.

To address RQ2, How are different levels of integrative thinking manifested in students’ written explanations of engineering design-based physics problems?, we examined the frequency and depth of IT connections across high-, mid-, and low-performing teams and used qualitative thick descriptions to illustrate how these connections were manifested within students’ written explanations. Integrative thinking was evident across all performance teams. High-performing teams demonstrated the greatest average number of IT connections, followed by low- and mid-performing teams; however, high- and low-performing teams demonstrated the same average number of substantial connections. Thus, the frequency and depth of integrative thinking did not vary monotonically with academic performance. Across performance teams, SBT–DBT and SBT–MBT were among the most frequently observed connections and were often substantial. Connections involving Computational Thinking were less frequent overall, although SBT–CT and MBT–CT connections were predominantly substantial when they occurred. Metacognitive connections were also evident across teams, but their depth varied. These qualitative analyses further demonstrated that different levels of integrative thinking were manifested through differences in how students connected multiple ways of thinking to explain, justify, and evaluate ideas and design decisions, rather than simply through the presence of multiple ways of thinking within the same explanation.

Taken together, the findings demonstrate that integrative thinking in engineering design-based physics is characterized by both the presence and depth of connections among ways of thinking. The quantitative findings show that students across different levels of academic performance can demonstrate integrative thinking, including substantial integration, while the qualitative findings illustrate how the depth of these integrations becomes visible in students’ written explanations. The combination of the WoT4EDP framework and the three-level analytic rubric therefore provides a methodological approach for examining integrative thinking as a multidimensional construct in engineering design-based physics contexts.

The primary contribution of this study is the operationalization of integrative thinking through an analytic framework that captures differences in the depth of integration among multiple ways of thinking. By distinguishing superficial, intermediate, and substantial integration and examining how these levels are manifested in students’ written explanations, this study provides a systematic way to investigate integrative thinking within authentic engineering design-based physics tasks. The findings further demonstrate that examining both the frequency and depth of students’ integration of multiple ways of thinking can reveal aspects of their thinking that may not be fully captured by traditional measures of academic performance.

The findings also have important instructional implications. The relatively limited integration of computational thinking and metacognitive reflection suggests opportunities to design instructional supports that more intentionally foster these forms of thinking and their integration with scientific, mathematical, and design reasoning. More broadly, this work demonstrates the value of assessing students’ reasoning through authentic engineering design tasks, offering insights into learning that extend beyond conventional measures of academic performance.

\section{ACKNOWLEDGMENTS}
ChatGPT 5.0 was employed by the lead author to wordsmith passages during the preparation of this manuscript. The lead author gratefully acknowledges Dr. Winter Allen and Lindsay Dawson for their valuable contributions to the consensus coding process.This work is supported in part by U.S. National Science Foundation grant 2021389. Opinions expressed are of the authors and not of the Foundation. 
\clearpage
\bibliography{references}
\end{document}